\documentclass[useAMS,usenatbib]{mn2e}
\usepackage{graphicx,subfigure}
\usepackage{amsmath}
\usepackage{rotating}
\usepackage{color}
\usepackage{times}

\title{A Genetic Approach to the History of the Magellanic Clouds}
\author[Guglielmo, M., Lewis G. F., Bland-Hawthorn, J.]{Magda Guglielmo\thanks{E-mail: m.guglielmo@physics.usyd.edu.au}, Geraint F. Lewis, Joss Bland-Hawthorn\\
Sydney Institute for Astronomy, School of Physics, A28, The University of Sydney, NSW 2006, Australia}
\begin{document}
\def \apj {ApJ}
\def \apss {Ap{\&}SS}
\def \mnras {MNRAS}
\def \aj {AJ}
\def \araa {ARA\&A}
\def \pasa {Publications of the Astronomical Society of Australia}
\def \aap {A\&A}
\def \aaps {A\&AS}
\def \apjl {ApJL}
\def \apjs {ApJS}
\def \pasj {PASJ}
\date{Accepted 2014 July 29.  Received 2014 July 21; in original form 2014 February 7}
\pagerange{\pageref{firstpage}--\pageref{lastpage}} \pubyear{2002}
\maketitle

\label{firstpage}
\begin{abstract}
The history of the Magellanic Clouds is investigated using N-body hydrodynamic 
simulations where the initial conditions are set by a genetic algorithm.
 This technique allows us to identify possible orbits for the Magellanic Clouds around the Milky Way, by directly comparing the simulations with observational constraints. We explore the parameter space of the interaction between the Magellanic Clouds and the Milky Way, considering as free parameters the proper motions of the Magellanic Clouds, the virial mass and the concentration parameter ($c$) of the Galactic dark matter halo. The best orbital scenarios presented here are considered with two different sets of parameters for the Milky Way disc and bulge components. The total circular velocity at the Sun's position ($\rm{R_{\odot}}=8.5\,\rm {kpc}$) is directly calculated from the rotation curve of the corresponding Galactic mass model. 
Our analysis suggests that the Magellanic Clouds have orbited inside the virial radius of the Milky Way for at least $3\,\rm{Gyr}$, even for low mass haloes. However, this is possible only with high values for the concentration parameter ($c\ge20$).
 In both orbital models presented here, the mutual interaction between the Magellanic Clouds is able to reproduce the observed features of the Magellanic System. 
\end{abstract}
%%%
%%%INTRODUCTION
%%%
\section*{ Introduction}

Gas accretion plays a fundamental role in the evolution of galaxies. In a few billion years, star formation will exhaust the gas reservoir in most spiral galaxies \citep{1980ApJ...237..692L}. This also holds true for the Milky Way. With a gas mass of $5\times10^{9}\,\rm{M_{\odot}}$ and a constant star formation rate of 1-3 M$_{\odot}$ year$^{-1}$, our galaxy is expected to run out of star formation fuel without an ongoing source \citep{ IAU:7916176,2012ARA&A..50..491P}. Moreover, the constant star formation rate inferred by the chemical abundance and kinematic of the solar neighbourhood, strongly suggests that the gas has been accreted from the intergalactic medium (either via the cold or hot accretion mode). However, how gas gets into galaxies remains a mystery. The most recent simulations present a complex picture in terms of (cold or warm) gas flows along channels  directed into the outskirts of galaxies. These filaments penetrate deep inside the halo and connect to the galactic centre from multiple directions \citep[e.g][]{2012MNRAS.423.3616D,2013MNRAS.436.3031V}.

 In a recent paper, \cite{2014arXiv1404.5514F} demonstrate that the nearby Magellanic System is sufficient to account for the star formation rate of the Milky Way for the next few billion years. In this instance, gas accretion is associated with infalling dark matter halos. Thus, satellite interactions can be an important mechanisms by which the Milky Way acquires gas. This may not be the main mechanism for other L$_{*}$ galaxies, since the Magellanic Clouds are a peculiar feature of the Local Group \citep{2011MNRAS.411..495J,2012MNRAS.424.1448R}, but the study of this system can reveal clues on the formation history of the Milky Way.

Large scale mapping of the 21 cm emission reveals the presence of several gaseous structures, evidence of an ongoing interaction between the Magellanic Clouds themselves and the Milky Way \citep{2003ApJ...586..170P,2005A&A...432...45B,2010ApJ...723.1618N}. The most prominent of these gas structures is the Magellanic Stream which extends $\sim140^{\circ}$ across the sky, passing through the South Galactic pole \citep{2010ApJ...723.1618N}. The two Clouds are linked together by the Magellanic Bridge, which contains almost 40$\%$ of the neutral gas mass in the Magellanic System \citep{2005A&A...432...45B}. Therefore, studying the origin  of these structures aids in understanding the underlying physical processes in galaxy interaction.

Over the past four decades, many groups have attempted to explain the presence of gas structures which characterise the Magellanic System. This difficult challenge is widely recognized as a benchmark for testing the usefulness of galaxy simulations \citep{2007ApJ...670L.109B}.
Formerly, the formation of these structures was explained as the result of multiple encounters between these satellites and the Milky Way, after which the gas within the Clouds was stripped by tidal forces \citep{1977MNRAS.181...59L,1994MNRAS.266..567G} or ram pressure \citep{1994MNRAS.270..209M,2005MNRAS.363..509M}. However, the new proper motion measurements show that the Clouds are on more energetic orbits, making the possibility of multiple encounters with the Galaxy more unlikely. Using proper motion measurements from the Hubble Space Telescope, \cite{2007ApJ...668..949B} show that the Large Magellanic Cloud (hereafter LMC) is more likely on its first passage around the Milky Way. Later, they extended this analysis to include the Small Magellanic Cloud (SMC), in order to explain the formation of the Magellanic Stream and the Leading Arm \citep{2010ApJ...721L..97B,2012MNRAS.421.2109B}. According to the first infall scenario, the Magellanic Stream and the Leading Arm are formed by interactions between the Clouds, while the Milky Way plays a secondary role. 
The model requires that the Magellanic Clouds formed a bound pair for at least 5 Gyr before falling into the Milky Way potential. Although the idea of a falling group is supported by previous models \citep{1980PASJ...32..581M,2011ApJ...742..110N}, it remains unclear whether  the Clouds were born as a bound pair at the epoch of galaxy formation or they have became a close pair only recently \citep{2005MNRAS.356..680B,2012ApJ...750...36D}

The main difficulty in modelling the orbit of the Magellanic Clouds is the large number of parameters required. Not only are the parameters directly related to the Clouds themselves uncertain, (such as their mass or their proper motion), but also those of the Milky Way potential, particularly the mass of the dark matter, its extension and its concentration parameters. There is considerable disagreement in the existing literature about the value of the Milky Way virial mass. Studies of the kinematics of the stellar halo show results ranging between $0.5-1.2\times10^{12}\,\rm{M_{\odot}}$ \citep{2005MNRAS.364..433B, 2012MNRAS.424L..44D, 2014Kafle}. On the other hand, results based on the radial velocity of the Milky satellite extend the upper limit by a factor of 2, suggesting a heavier halo with mass with a mass of $2.0-3.0\times10^{12}\,\rm{M_{\odot}}$ \citep{2011MNRAS.414.1560B, 2013ApJ...768..140B}. In addition, the mass-concentration relation sets the concentration around 10-17, for haloes with mass of the order of 10$^{12}\,{\rm{M_{\odot}}}$  \citep{2002ApJ...573..597K, 2008MNRAS.391.1940M}. This range is defined by using dark matter only simulations, but the presence of the baryonic component (stars and gas) can cause the compression of the dark matter halo. However, studies of the kinematics of the Milky Way stellar halo, show that the concentration parameter can be higher ($18\le{c}\le24$) than the values predicted by simulations \citep{2005MNRAS.364..433B, 2012MNRAS.424L..44D, 2014Kafle}. Therefore, the uncertainties on the Milky Way mass pose a fundamental limitation in understanding the orbit of the Magellanic Clouds.
In particular, different combinations of the virial mass and concentration parameters lead to different values of the circular velocity of the Milky Way at the position of the Sun. This parameter has a strong influence on the orbits of the Magellanic Clouds, since it has been shown that higher values than the standard IAU value ($220\,\rm{km\,s^{-1}}$) increase the number of close encounters between LMC and the Galaxy \citep{2009MNRAS.392L..21S,2010ApJ...725..369R,2012ApJ...750...36D,2012ApJ...759...99Z}. .

Several approaches have been proposed to study the parameter space of the Magellanic Clouds. \cite{2010ApJ...725..369R} use a genetic algorithm combined with a restricted N-body integration scheme \citep{1972ApJ...178..623T,1999RvMA...12..309T} to address the formation of the stream. \cite{2012ApJ...750...36D} explore a wide range of orbital models and use a multi-component N-body representation for only the SMC, in order to investigate the tidal effects on its disc and the formation of the main structures in the system. However, due to the particular integration scheme used, these studies do not fully model the interaction between the Clouds themselves. \cite{2012MNRAS.421.2109B} consider both LMC and SMC as multi-component N-body systems, but the high resolution of their simulation does not allow for a complete parameter search of all the possible orbital configurations.

In this paper, we present a genetic algorithm combined with full N-body Gadget2 simulations \citep{2005MNRAS.364.1105S}. The aim is to address the formation of the Magellanic System with a direct comparison between the outcome of simulations and the observed properties of the system. The orbits are selected with the only constraints being the two encounters that occurred between the Clouds occured in the last 3 Gyr. This is a common feature in previous models of the Magellanic Clouds' orbit, both in the first infall scenario \citep{2012MNRAS.421.2109B} and in more traditional scenarios \citep{2010ApJ...725..369R,2012ApJ...750...36D}.

Traces of these encounters exist in the recent star formation history (SFH) of the Clouds, as strong tidal interactions between LMC and SMC might have caused episodic star bursts in both galaxies \citep{2005MNRAS.356..680B,2005MNRAS.358.1215P}. Studies of the SFH show that it has been increasing in the last few Gyr, corroborating the hypothesis of strong interactions between the Clouds \citep{2005AJ....130.1083M,2009ApJ...705.1260N}. \cite{2009AJ....138.1243H} find two common peaks in SFH in both Clouds, one $2\rm{-}3\,\rm{Gyr}$ ago and again $400\,\rm{Myr}$ ago, suggesting the time of the interactions. 

This paper is organized as follows: in \S\ref{sec:GA} we present an
introduction to the genetic algorithm and its application to the Magellanic Clouds problem; in \S\ref{sec:NumMod} and \S\ref{sec:PS}, we describe the numerical model for the Magellanic Clouds and the Milky Way and their parameter space; and in \S\ref{sec:MF} we focus on how the best solution is selected. As described in \S\ref{sec:PS}, the parameters related to the dark matter halo (its mass and concentration) are free to span in the range given by the observational and theoretical constraints, while the parameters for the Milky Way disc and bulge are fixed. The selected orbits are described in \S\ref{sec:BestOrbi}. 
%---------------------------------
%------GENETIC ALGORITHM-----------
%--------------------------------
\section{Why do we need a Genetic Algorithm?}
\label{sec:GA}
The study of galaxy interactions requires a complete knowledge of the parameters which lead to the
observed configuration. The difficulty is that in the case of orbital integration, the parameter space is very large. Even just considering the simplest case of two galaxies, it is crucial to know their present day positions and velocities, total mass and mass distributions. Adding a third body, such as a central galaxy, increases the number of parameters involved by at least 25 per cent. Therefore, dealing with the problem of multi-body interactions means coping with a higher dimensional space, normally too high for standard approaches, such as Monte Carlo chains.

Emulating the biological concept of evolution, the genetic algorithm (GA) is a powerful tool to explore a complex parameter space. In biology, given a set of possible genetic sequences (``population of individuals''), the fittest organisms are those strong enough to survive and reproduce themselves in their environments: nature selects creatures with a high probability of survival (``survival of the fittest''). In optimisation problems, given a set of ``possible solutions'', the best is the one which better adapts to the requirements imposed by the model. The genetic algorithm mimics the reproduction, mutation and selection to arrive at the fittest set of parameters.
 
Keeping the same terminology from the biological world, a \emph{gene} is the value of a particular parameter and the \emph{phenotype} encodes the collection of all parameters which describe a possible solution. When all the phenotypes are created, they are sorted according to their value of the merit function. A simple genetic algorithm consists of the following steps \citep{1995ApJS..101..309C}:
\begin{enumerate}
\item Start by randomly generating an initial population of phenotypes, each representing a possible solution.
\item \label{it:2} Evaluate the fitness of each member of the current population.
\item Select a pair of genotypes (``parents'') from the current population and breed them, based on their merit. In this way, two new solutions are generated (``offspring''). Repeat this step until the number of offspring produced equals the number of individuals in the current population.
\item Replace the old population with the new one.
\item Repeat from step \ref{it:2} until the fitness criterion is satisfied.
\end{enumerate}
The evolution process is driven by different types of operators. The first one is \emph{elitism}: the fittest member of the current population is cloned over the next generation. This guarantees that the maximum values of the fitness function can never fall. The \emph{breeding process} will depend on the selection of the parents: the individuals with highest fitness have higher probability to be selected. A random portion of the parent genome is mixed with the other phenotype; this process is named \emph{crossover}. Whether or not the offspring is generated will depend on a pre-defined probability (\emph{crossover rate}). The final operation is the \emph{mutation}, which mimics the probability that a particular gene can mutate in the next generation. In the case of a genetic algorithm, the mutation will flip a bit in the phenotype, according to some probabilities determined by the \emph{mutation rate}.

Because of its versatility, the genetic algorithm can be applied to different problems which require the exploration of a large parameter space. Pioneering work applied this algorithm to different astrophysical problems, such as fitting the light and rotation curves of galaxies \citep{1995ApJS..101..309C}, determining orbital parameters of interacting galaxies \citep{1998A&AS..132..417W}, designing filter systems \citep{1998MNRAS.299..176O} and inverting gravitational lensed images \citep{2005PASA...22..128B}.

Using the same recipe described above, we present a genetic algorithm which selects the best individuals based on the results of N-body simulations. Previous applications of genetic algorithms by \cite{2007A&A...461..155R,2009ApJ...691.1807R,2010ApJ...725..369R} used a restricted N-Body simulation to model the interaction between the Clouds. The restricted N-body is computationally less expensive than a full N-body simulation. However, this model does not consider any mass loss and therefore the masses of the Clouds are considered constant in time. In reality, due to the interaction between each other and the Milky Way, the masses of the two Clouds evolve, influencing their orbits. In this paper, we extend the \cite{2010ApJ...725..369R} analysis, by using the Gadget2 simulation code \citep{2005MNRAS.364.1105S}.

In order to identify the past positions and velocities of both clouds, we perform a three body orbit integration backward in time, starting from today back to 3 Gyr ago \citep{1980PASJ...32..581M}. Then, the forward integration starts, using Gadget2 code which has been modified in order to include a static Milky Way potential, as described in the next section. We integrate over a period of $3\,\rm{Gyr}$ and the last Gadget snapshot (corresponding to the present day configuration) is used to assign the fitness function and, therefore, to select the best individuals.
%____________________________________________________________
%_____________________Parameter N-Body_________________
%____________________________________________________________________
\begin{table}
\centering
\begin{tabular}{c c c}
\multicolumn{1}{c}{ } &
\multicolumn{1}{c}{\textbf{LMC}}&
\multicolumn{1}{c}{\textbf{SMC}}\\
\hline
\hline
$\rm{M}_{\rm{halo}}\,\rm{(\rm{M}_{\odot})}$& $2.13\times10^{10}$ &$0.50\times10^{10}$\\ $\rm{M}^*\,\rm{(\rm{M}_{\odot})}$ &$0.31\times10^{10}$& $0.12\times10^{10}$\\
$\rm{r_{disc}}\;{(\rm{kpc})}$&$1.4$&$1.25$\\
$\rm{h_{disc}}\;{(\rm{kpc})}$&$0.3$&$0.2$\\
$\rm{r}_{\rm{halo}}\;{(\rm{kpc})}$& $10$ & $5.0$\\
\hline
\hline
\end{tabular}
\caption{Initial conditions for LMC and SMC.}
\label{tab:N-body}
\end{table}
%____________________________________________________________
%___________________________________________________________________
%____________________________________________________________________

\section{Numerical Model}
\label{sec:NumMod}
In order to simulate the evolution of the Magellanic Clouds around the Milky Way, we model the latter as a static multicomponent potential, consisting of a disc, a central bulge and a dark matter halo. 
 The disc is assumed to be a Miyamoto-Nagai potential \citep{1975PASJ...27..533M}
\begin{equation}
\Phi_{\rm{disc}}(R,z)=-\frac{\rm{GM_{disc}}}{\left(R^2+\left(r_{\rm{disc}}+{\sqrt{(z^2+b^2)}}\right)^2\right)^{1/2}},
\end{equation}
while the bulge component follows a Hernquist profile \citep{1990ApJ...356..359H}
\begin{equation}
\Phi_{\rm{bulge}}(r)=-\frac{\rm{GM_{{bulge}}}}{r_{\rm{bulge}}+r}\mbox{.}
\end{equation}
The dark matter halo is given by a Navarro, Frenk and White \citep{1997ApJ...490..493N} (hereafter, NFW) potential
\begin{equation}
\Phi_{\rm{halo}}(r)=-\frac{\rm{GM_{\rm{halo}}}}{r}\ln \left(\frac{r}{r_{\rm{halo}}}+1\right).
\label{eq:Halo}
\end{equation}  
The halo mass $\rm{M_{halo}}$ scale is related to the virial mass via 
\begin{equation}
\rm{M_{halo}}=\frac{\rm{M_{vir}}}{\ln{(c+1)-c/(c+1)}},
\label{eq:Mhalo}
\end{equation}
and the halo radius $r_{\rm{halo}}$ is related to the virial radius 
\begin{equation}
r_{\rm{halo}}=\frac{\rm{R_{\rm{vir}}}}{c},
\label{eq:rhalo}
\end{equation}
where $c$ is the concentration parameter.
 As we consider a low mass profile for both Clouds, the effects of the dynamical friction between the Clouds and the Milky Way halo are not explicitly considered in the equation of motion; while the dynamical friction between the two Clouds is accounted by modelling the two Clouds with a live dark matter halo.

The initial conditions for the Magellanic Clouds are generated using GalactICS \citep{2008ApJ...679.1239W}. Both galaxies are modelled as truncated dark matter halos and an exponential disc component (see \cite{2008ApJ...679.1239W} for further detail on the GalactICS code).

\begin{figure*}
\centering
\includegraphics[scale=0.8]{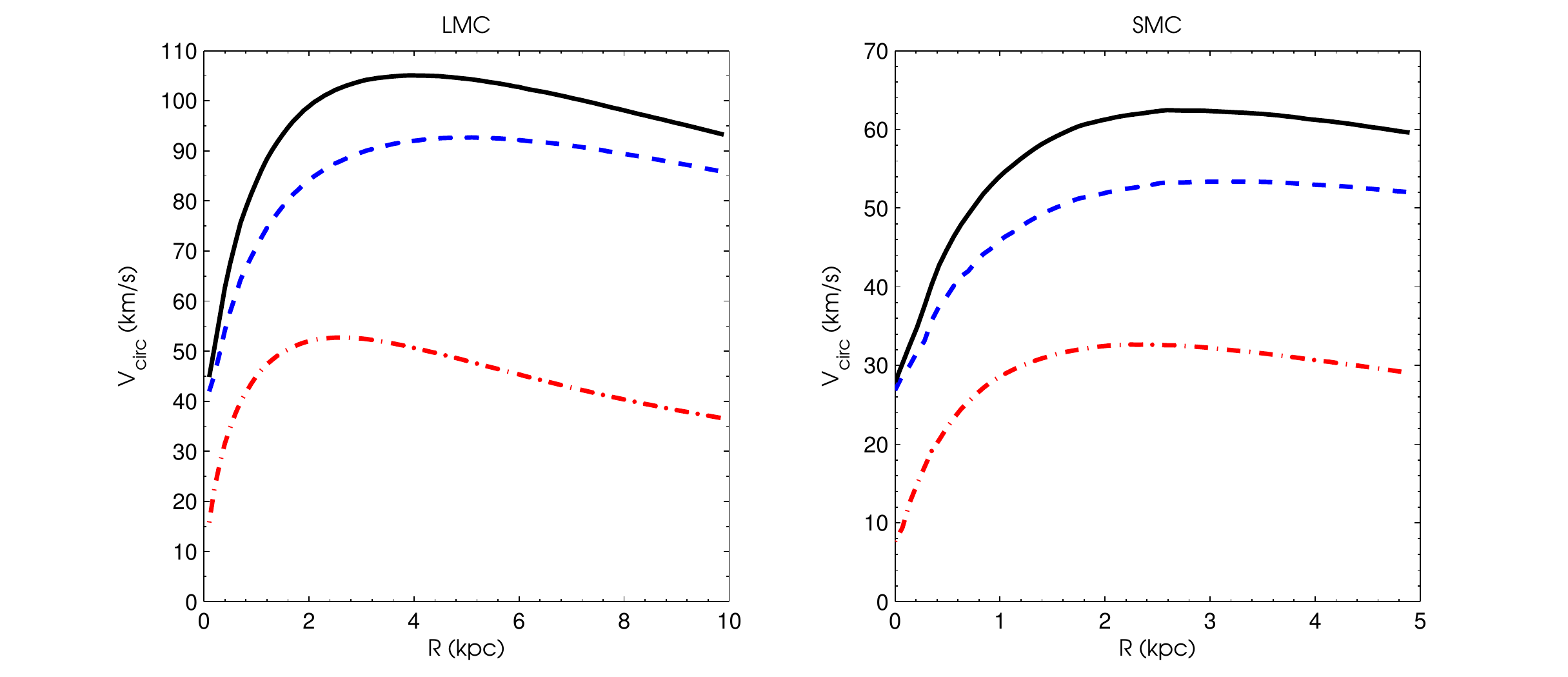}
\caption{  The initial rotation curves for LMC (left) and SMC (right) are plotted. For each Cloud, the contribution of the halo and the initial baryonic component (star+gas) are plotted in blue and red respectively, while the solid black line indicates the total rotation curve.}
\label{fig:MCcurve}
\end{figure*}

 The resulting initial rotation curves are shown in figure \ref{fig:MCcurve}. As shown in the left panel of Figure \ref{fig:MCcurve}, the total circular velocity of the initial LMC peaks at 105 km/s (black solid line), which is consistent with the recent estimation by \cite{2014ApJ...781..121V} who quote the rotation curve of the LMC to be $91\pm18\;\rm{km\,s^{-1}}$ \citep{2014ApJ...781..121V}. The initial mass of LMC is chosen so that the total mass within 9 kpc is 19$\times10^9$ M$_{\odot}$; whereas the right panel shows the rotation curve of the initial SMC. The total rotation curve peaks around 64 km/s, with the total mass within 3 kpc equal to 2.8 $\times10^9$ M$_{\odot}$ \citep{2004ApJ...604..176S}.

During the initial parameter search, we did not include any gas component for the Clouds. This is because a hydrodynamical simulation will increase the time of a single Gadget run. However, for the best set of parameters found by the GA, a gaseous disc was added to both of the Magellanic Clouds (see section \S \ref{sec:BeslaSolution}).

In order to save computational time, the total number of particles used for each run is $~10^4$, but for the final results this number increases by a factor of 10. In each galaxy, the number of particles is chosen so that the mass of each particle is roughly the same, respecting the total mass ratio between the Clouds (1:10). Before adding the external potential of the MW, each modelled galaxy was simulated in isolation in order to test the stability of the system.
%------------------------------------------
%------------Rot Curve MCs------------------
%------------------------------------------
%------------------------------------------
%-------------------------------------------
%------------------------------------------
\section{The parameter space}
\label{sec:PS}
The genetic algorithm considers in total six independent parameters; the virial mass and the concentration of the Milky Way halo and the west and north proper motion components for both LMC and SMC. Table \ref{tab:GAparam} lists the parameters and their range. These parameters will change both the equation of motion and the present day velocities of the Clouds. In Table \ref{tab:GAparam}, the two different models referred to two different sets of parameters used for the Milky Way disc and bulge. The adopted parameters for these components are listed in Table \ref{tab:Disk-Bulge}.
In this section, we describe in greater detail the parameters used for the Milky Way potential and the proper motion of the Clouds. 

\subsection{The Milky Way Parameters}
\label{sec:MW}
The orbital history of the Magellanic Clouds strongly depends on the potential of the Milky Way, in particular the mass of the dark matter halo. This dependency is not only related to the orbit of the Clouds around their host, but also the evolution of the SMC around LMC can change dramatically for a different choice of the Milky Way halo mass.
Due to our position within the Milky Way, it is hard to directly estimate the value of the halo mass.
The kinematics of Blue Horizontal Branch stars in the Galactic halo suggest that the virial mass of the Milky Way is around $10^{12}\,\rm{M_{\odot}}$ \citep{2008ApJ...684.1143X,2012ApJ...761...98K}. Combining the new proper motion measurements of Leo I \citep{2013ApJ...768..139S} with numerical simulations of Milky Way-size dark matter halo, \cite{2013ApJ...768..140B} constrain the mass of the Milky Way's dark matter halo at $1.6\times10^{12}\,\rm{M_{\odot}}$ with 90 per cent confidence in the range $[1.0,2.4]\times10^{12}\,\rm{M_{\odot}}$. In the following analysis, this parameter is assumed to vary between $0.90\times10^{12}\,\rm{M_{\odot}}$ and $2.0\times10^{12}\,\rm{M_{\odot}}$.

The dark matter profile in equation \ref{eq:Halo} depends on the virial radius and concentration parameter, as well as the virial mass. The virial radius $\rm{R}_{\rm{vir}}$ is calculated for each value of the virial mass, using the equation
\begin{equation}
\label{eq:rvir}
\rm{R}_{\mathrm{vir}}=\left(\frac{2M_{\mathrm{vir}}G}{H^{2}_0\Omega_{m}\Delta_{th}}\right)^{1/3},
\end{equation}
where $H_0=70.4\;{\mathrm{kms}}^{-1}\,{\mathrm{Mpc}^{-1}}$, $\Omega_{m}=0.3$ and $\Delta_{th}=340$.
The concentration $c$ is considered a free parameter in the range 1 to 30. In previous work the concentration of the halo is fixed at a standard value of 12 \citep{2012MNRAS.421.2109B,2012ApJ...750...36D}. By fitting the kinematics of the stellar halo of our Galaxy, recent work by \cite{2014Kafle} claims that there is the possibility of a more concentrated halo. Therefore, allowing the concentration to vary will allow us to study the consequences of such a model on the orbit of the Clouds.

The circular velocity of the Milky Way at the position of the Sun V$_{\mathrm{cir}}$ influences the orbital history of the Clouds \citep{2009MNRAS.392L..21S, 2010ApJ...725..369R}. This is because proper motions are measured relative to the Solar System, so the rotational velocity of the Sun is needed to convert the velocities of the Clouds to a Galactocentric frame (see \S\ref{ProperMotion}).
Although the IAU standard value for $V_{\rm{cir}}$ is $220\;\rm{km\,s^{-1}}$, recent estimations \citep{2009ApJ...700..137R,2011MNRAS.414.2446M} infer higher values for this parameter. In this work, the circular velocity is directly calculated from the rotation curve of the Milky Way.

 Although the disc and the bulge parameters of the Milky Way are not considered as free parameters, here two different sets of parameters are used to investigate how this choice can influence the orbit of the Clouds (see Table \ref{tab:Disk-Bulge}). The potential of the disc and bulge will influence the value of the rotational velocity of the Milky Way.
%-------------------------------
%			PM
%-----------------------------
\begin{table*}
\centering
\begin{tabular}{|c c c c|}
\hline
& LMC & SMC & References\\
\hline
$m-M$&$18.50\pm0.1$&$18.95\pm0.1$&\cite{2002AJ....124.2639V}, \cite{2000AA...359..601C}\\
$V_{\mathrm{sys}}\,\rm{(km\,s^{-1})}$& $262.2$&$146.0$&\cite{2002AJ....124.2639V}, \cite{2006AJ....131.2514H}\\
$(\alpha,\delta)\,$(deg)& $(81.9,-69,9)$&$(13.2,-72.5)$&\cite{2002AJ....124.2639V}, Smith et al (2007)\\
$(l,b)\,$(deg)&$(280.253,-32.5)$& $(301.5,-44.7)$&-\\
 \hline
\hline
\end{tabular}
\caption{Adopted values for the distance moduli, systemic velocity and Galactic Coordinates for both Clouds. These values are used for converting the velocity in the Galactic frame in equation \ref{eq:val}.}
\label{tab:velocityparam}
\end{table*}
%%---------------------------------------------------------------------------
%---------------------------------------------------------------------------
%------------------------Proper Motion----------------------------------
%---------------------------------------------------------------------------
\subsection{The Proper Motion of the Clouds}
\label{ProperMotion}
A crucial step in studying the orbital evolution of galaxies is the choice of their present day velocities. In the last decade, several proper motion catalogues for the Magellanic Clouds have been published \citep{2006ApJ...652.1213K,2009AJ....137.4339C,2013ApJ...764..161K,2010AJ....140.1934V}. Even though they were obtained with different techniques, they are largely consistent with each other. The main problem is that by using different values, the orbital history of the Clouds changes completely.

With the aim of selecting the tangential velocities in a more general way, we allow the proper motions to span within the error range of the catalogue presented in \cite{2010AJ....140.1934V}. This catalogue summarizes the results of CCD and photographic observation, using a ground based telescope over a baseline of 40 years. The advantage of using this catalogue is to include, within the error, other proper motion measurements.
 
Once the proper motions in both directions are selected, the velocities of the Clouds need to be corrected with respect to the position and velocity of the Sun 
\begin{equation}
\mathbf{r_{\odot}}=(-\rm{R}_{\odot},0,0), \qquad \mathbf{v_{\odot}}=(U_{\odot},V_{\odot}+V_{\mathrm{cir}},W_{\odot}),
\end{equation}
where $(U_{\odot}, V_{\odot}, W_{\odot})=(11.1, 12.24, 7.25)\,\rm{kms}^{-1}$ \citep{2010MNRAS.403.1829S} and the distance of the Sun from the Galactic Centre is fixed at $\rm{R}_{\odot}=8.5\,\rm{kpc}$ .

The proper motions in the direction of west and north are defined as 
\begin{equation}
\mu_W=-\cos{\delta}\frac{d\alpha}{dt}, \qquad \mu_N=\frac{d\delta}{dt},
\end{equation}
For each value of the proper motion selected by the GA and the total circular velocity corresponding to the selected Milky Way's model, the present day velocities are transformed into the Galactocentre frame. Following \cite{2002AJ....124.2639V}, we adopt a Cartesian coordinate system with the origin at the Galactic Centre: the $z\mathrm{-axis}$ pointing toward the Galactic north pole, the $x\mathrm{-axis}$ pointing from the Sun to the Galactic Centre; and $y\mathrm{-axis}$ aligned in the direction of the Sun's Galactic rotation. In this system, the velocities are calculated by 
\begin{equation}
v^{i}=v^{i}_{\odot}+V_{\mathrm{sys}}u^{i}_{0}+D\mu_{W}u^{i}_{1}+D\mu_{N}u^{i}_{2},
\label{eq:val}
\end{equation}
where $V_{\rm{sys}}$ is the line-of-sight systemic velocity and D is the distance to the Galaxy, given in Table \ref{tab:velocityparam}.
The vectors $\mathbf{u_0}$, $\mathbf{u_1}$ and $\mathbf{u_2}$ are the unit vectors from the Sun in the direction of the Clouds, given by
\begin{align*}
\mathbf{u}_{0}&=(\cos{l}\cos{b},\sin{l}\cos{b},\sin{b})\\
\mathbf{u}_{1}&=-\frac{1}{\cos{\delta}}\frac{\partial{\mathbf{u}_0}}{\partial{\alpha}}\\
\mathbf{u}_{2}&=\frac{\partial{\mathbf{u}_0}}{\partial{\delta}}
\end{align*}
where the Galactic Coordinates $(\ell,b)$ are those listed in Table \ref{tab:velocityparam}.

%----------------------------------------
%---- TABLE PARAMTERS/RESULTS GA
\begin{table*}
\centering
\begin{tabular}{ccccc}
\hline
\hline
Parameters&Range&Model1&Model2&References\\
\hline
\hline
$\rm{M_{vir}}\;(10^{12}\; \rm{M}_{\odot})$&$[0.90,2]$&$1.00$&$1.27$&\cite{2008ApJ...684.1143X,2012ApJ...761...98K}\\
&&&\\
$\rm{c}$ &$[1,\;30]$&$27.3$&$20.5$&\cite{2005MNRAS.364..433B,2012MNRAS.424L..44D}\\
&&&&\cite{2013ApJ...773L..32R},Kafle et al (2014)\\
&&&\\
$(\mu_{\rm{W}},\,\mu_{\rm{N}})_{\rm{LMC}}\; (\rm{mas/yr})$ &$(-1.89\pm0.27,\,0.39\pm0.27)$&
$(-1.87,\,0.38)$&$(-2.03,\,0.19)$&\cite{2010AJ....140.1934V}\\
&&&\\
$(\mu_{\rm{W}},\,\mu_{\rm{N}})_{\rm{SMC}}\; (\rm{mas/yr})$ &$(-0.98\pm0.30,\,-1.10\pm0.29)$&$(-1.08,\,-1.04)$&$(-0.98,\,-1.20)$&\cite{2010AJ....140.1934V}\\
&&&\\
$\Omega\,(\rm{km\,s^{-1}kpc^{-1}})$&$[28.0,\,32.0]$&$30.3$&$30.3$&\cite{2010MNRAS.402..934M}\\
&&&\\
$\rm{V_{cir}}\;\rm{(km\,s^{-1})}$&-&$245.3$&$245.8$&\cite{2011MNRAS.414.2446M}\\
\hline
\hline
\end{tabular}
\caption{Parameter range and genetic algorithm best values. The first and second columns describe the parameters used in the genetic algorithm with their range. Note that the circular velocity is not a free parameter, but is calculated from the rotation curve, therefore it depends on the particular choice of the virial mass and concentration. The following columns describe the results for the two different models of the disc and bulge used in this work (see Table \ref{tab:Disk-Bulge}). The last column shows the references for each parameter value.}
\label{tab:GAparam}
\end{table*}

\section{The merit Function }
\label{sec:MF}
The use of the genetic algorithm, together with the N-Body integration, allows an automatic search in the parameter space, with a simultaneous comparison between model and observations. The choice of the merit function is the most critical step of the algorithm; the wrong function can lead to  the convergence of the algorithm to the wrong solution. Following \cite{2010ApJ...725..369R}, the total function $F$ is defined in such a way that it will return a number in the range 0 and 1, according to the ability of the single individual to satisfy the imposed requirements. Here, $F$ is chosen to be the product of three different functions, each representing a particular requirement
 \begin{equation}
 F=f_{1}*f_{2}*f_{3}.
 \label{eq:TotFit}
\end{equation} 

The first condition is that the final position and velocity of the main body of each galaxy is consistent with the observed values for the Magellanic Clouds. During the N-body simulations, the formation of particle structures could cause a deviation of the centre of mass orbit from the one calculated using the point mass approximation. This deviation can lead the Clouds to be in
the wrong position in the sky. In order to reproduce their present day positions and velocities, a comparison between the simulation results and the observed values is made through the equation 
\begin{equation}
 f_1=\prod_{i=1}^{12}\frac{1}{1+\left(\frac{x_i-x_{i_\mathrm{exp}}}{x_{i_\mathrm{exp}}}\right)^2},
 \label{eq:f1}
\end{equation} 
where $x_i$ indicates the (x, y, z) positions and the corresponding velocity components of the centre of mass of each Cloud, and $x_{i_\mathrm{exp}}$ is the corresponding observed values. To better compare these quantities, the centre of mass is calculated using only particles bound to the main bodies.

The second condition $f_2$ is on the orbit of SMC around LMC. As shown by \cite{2004AJ....127.1531H,2009AJ....138.1243H}, the star formation history of both Magellanic Clouds present two common peaks at $\rm{T}\sim2.5\,\rm{Gyr}$ and $\rm{T}\sim0.4\,\rm{Gyr}$. These can be interpreted as two close encounters between the Clouds.
With the aim of reproducing these features, the best parameters are defined in such a way that there have been at least two encounters between SMC and LMC 
 \begin{equation}
 f_2=\prod_{i=1}^2\frac{1}{1+\left(\frac{\rm{t_j}-\rm{T_j}}{\sigma}\right)^2}\quad \mbox{where}\;j=1,2
 \label{eq:f2}
\end{equation} 
where $\rm{t_1}$ ($\rm{t_2}$) is the time of the first (second) encounter; and $\rm{T}_1=2.5\,\rm{Gyr}$ ($\rm{T}_2=0.4\,\rm{Gyr}$) corresponds to the time of the peaks in the star formation of both Clouds \citep{2009AJ....138.1243H}. Although the values of $\rm{T}_1$ and $\rm{T}_2$ are well constrained, demanding two encounters at fixed time is a very strong condition and the algorithm might require a large number of generations to reach the convergence. In order to release this requirement, we set $\sigma$ at $0.5\,\rm{Gyr}$.

The last condition is related to the angular velocity at the position of the Sun ($\rm{R}_\odot=8.5\,\rm{kpc}$), defined as $\omega=\frac{(\rm{V}_\odot+\rm{V_{cir}})}{\rm{R}_{\odot}}$, where  the peculiar motion of the Sun is $\rm{V}_\odot=12.24\,\rm{km\,s^{-1}}$ (see \S\ref{ProperMotion}). Therefore, the $f_3$ condition in equation \ref{eq:TotFit} is given by
\begin{equation}
 f_3=\frac{1}{1+\left(\frac{{\omega}-\rm{\Omega_{exp}}}{\sigma}\right)^2},
 \label{eq:f2}
\end{equation} 
The values of $\Omega_{exp}$ and $\sigma$ are such that the angular velocity of each individual belongs to the range $[28.0,\,32.0]\,\rm{km\,s^{-1}kpc^{-1}}$, consistent with the range found by \cite{2010MNRAS.402..934M}.
%------------------------------------------------------------
%-------------------Parameter MW Disk---------------
%------------------------------------------------------------
\begin{table}
\centering
\begin{tabular}{c c }
Parameter & Value\\
\hline
\multicolumn{2}{c} {Model 1 }\\
\hline
$\rm{M_{disc}}\;(10^{10}\;\rm{M_{\odot}})$& 5.5\\
$\rm{r_{disc}}\;\;(\rm{kpc}) $& 3.5\\
$\rm{b_{disc}}\;\;(\rm{kpc}) $& $\rm{r_{disc}}/5$\\
$\rm{M_{bulge}}\;(10^{10}\;\rm{M_{\odot}})$& 1.0\\
$\rm{r_{bulge}}\;\;(\rm{kpc}) $& 0.7\\
\hline
\multicolumn{2}{c} {Model 2 }\\
\hline
$\rm{M_{disc}}\,(10^{10}\;\rm{M_{\odot}})$& 7.6\\
$\rm{r_{disc}}\,(\rm{kpc}) $& 6.5\\
$\rm{b_{disc}}\,(\rm{kpc}) $& $0.3$\\
$\rm{M_{bulge}}\,(10^{10}\;\rm{M_{\odot}})$&2.4\\
$\rm{r_{bulge}}\,(\rm{kpc}) $& 0.31\\
\end{tabular}
\caption{Disc and bulge parameters used for the two models of the Milky Way}\label{tab:Disk-Bulge}
\end{table}
%---------------------------------------------------------------------------
%---------------------------------------------------------------------------
%---------------------------------------------------------------------------
%-------------------The best orbital model---------------
%---------------------------------------------------------------------------
\begin{figure*}
\centering
\includegraphics{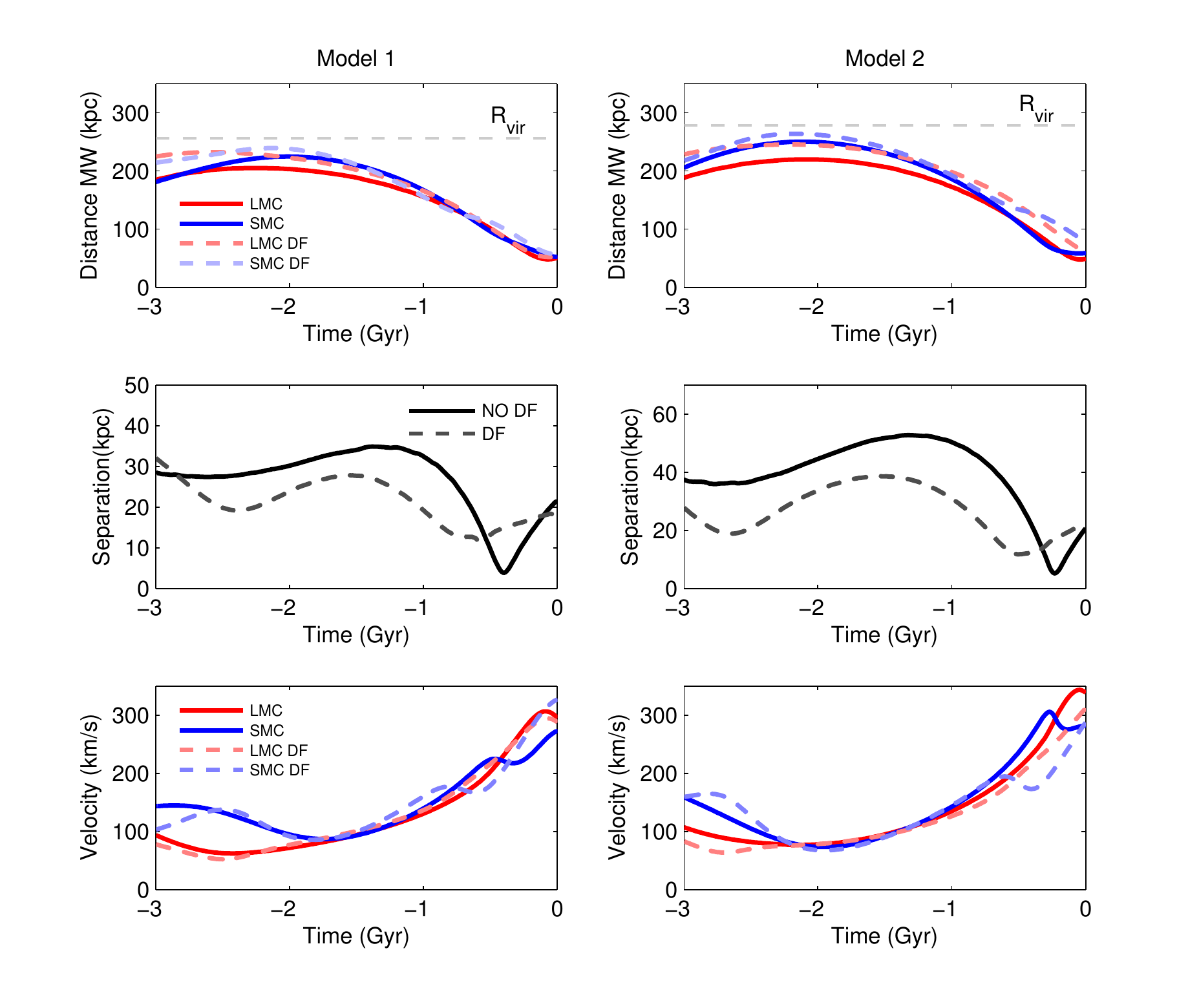}
\caption{Orbit for the best individuals in Model 1 (\emph{first column}) and Model 2 (\emph{second column}). The first row of the figure shows the orbits of both Clouds around the Milky Way. In both cases, the Clouds are orbiting within the virial radius (shown here as dashed gray line) of the Milky Way for the last $3\,\rm{Gyr}$. In the second row, the distance between LMC and SMC is plotted as a function of time. The last row shows the total velocity for both Clouds. In all panels, the dashed lines show the effect of the dynamical friction on the Clouds motion due to the Milky Way halo (see section \S\ref{sec:DF}).} 
\label{fig:OrbitInfo}
\end{figure*}
%---------------------------------------------------------------------------
%---------------------------------------------------------------------------
%
%Rot Curve Milky Way
%
\begin{figure*}
\centering
\includegraphics{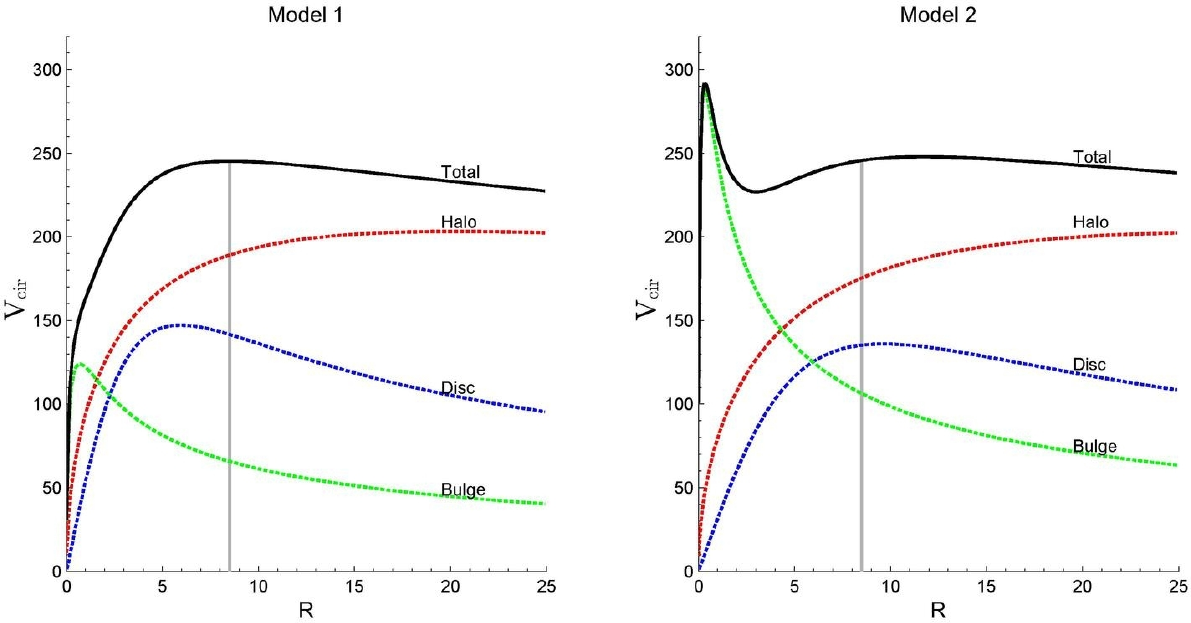}
\caption{Rotation curve of the Milky Way for Model 1 (\emph{left panel}) and Model 2 (\emph{right panel}), shown for the contribution of each component (NFW halo, Myamoto-Nagasai disc and spherical bulge ) and their total. For each model, the rotation curves due to the disc and bulge are fixed, while the halo contribution is chosen by the genetic algorithm. The adopted values for the virial mass of the halo are $0.99\times10^{12}\;\rm{M_{\odot}}$ for Model 1 with a concentration parameter of $27.3$; a mass of $1.27\times10^{12}\;\rm{M_{\odot}}$ and a concentration parameter equal to $20.5$ for Model 2. In both panels, the solid grey line indicates the position of the Sun ($\rm{R_{\odot}}=8.5\,\rm{kpc}$) and its intersection with the total curve provides the circular velocity adopted for each model ($\rm{V_{cir}}=245.3\,\rm{km\,s^{-1}}$ Model 1 , and $\rm{V_{cir}}=245.8\,\rm{km\,s^{-1}}$ Model 2).} 
\label{fig:RotCurve}
\end{figure*}
\section{The best orbital models}
\label{sec:BestOrbi}
The genetic algorithm studies the evolution of a first generation consisting of $50$ individuals randomly selected on a sample of possible solutions. The evolution ends when the maximum number of generations reaches $50$. The results presented here refer to the best individual (\emph{high fitness value}) in the last generation.

As mentioned in \S\ref{sec:NumMod}, for each run of the genetic algorithm, the Clouds are modelled with a small total number of particles and no gas particles are included. However, for the two best solutions presented here, a new simulation with Gadget2 is carried out, using the same mass model for the Clouds described in Table \ref{tab:N-body}, but with a total particle number of $3\times10^5$. In this final simulation, a gas component is added to each of the Clouds. It is important to note that the best solution found by the genetic algorithm strongly depends on the particular choice of the Clouds' mass distribution. Therefore, it is important that the total mass of the each Cloud remains the same as the one used in the parameter search. We add a disc of gas, with the mass defined in such a way that the gas fractions (ratio between the mass of the gas and the total baryonic mass) are $f_{gas}=0.3$ and $f_{gas}=0.7$ for LMC and SMC respectively \citep{2012MNRAS.421.2109B}.

All sky final gas distribution for Model 1 and Model 2 is shown on a zenithal equal area projection (ZEA) in the right and left panels (respectively) in Figure \ref{fig:Zea}. Both models are able to reproduce the location of the main features of the Magellanic Stream and Bridge. We present the final distribution of stars (disk particles) on the same projection in Figure \ref{fig:ZeaStar}. In general, there are only few stars in the location of the Magellanic Stream with total mass of $1.0\times10^6\,\rm{M_{\odot}}$. Our simulations show the presence of a bridge of stars connecting the two Clouds, formed after the last encounter.
\subsection{Model 1: An Initial Orbital Scenario}
\label{sec:BeslaSolution}
The parameter Milky Way disc and bulge models are those corresponding to the fiducial model used in \cite{2007ApJ...668..949B}. The Miyamoto-Nagai disc is chosen to have a mass of $\rm{M_{disc}}=5.5\times10^{10}\,\rm{M_{\odot}}$, a radius of $\rm{r_{disc}}=3.5\,\rm{kpc}$, and disc scale height given by $\rm{r_{disc}}/5.0\,\rm{kpc}$. The bulge has a mass of $\rm{M_{bulge}}=1.0\times10^{10}\,\rm{M_{\odot}}$ and a radius of $0.7\,\rm{kpc}$. As mentioned in \S\ref{sec:MW}, the virial radius and the concentration parameters vary in the range defined in Table \ref{tab:GAparam}. For the best solution, the virial mass is $0.99\times10^{12}\rm{M_{\odot}}$, the virial radius is $256\,\rm{kpc}$ and the concentration is equal to $27$. The final rotation curves for the Milky Way corresponding to this model are shown in the left panel of Figure \ref{fig:RotCurve}.
The proper motions for the best individual are:
\begin{subequations}
\begin{align}
(\mu_{\rm{W}},\mu_{\rm{N}})_{{\rm{LMC}}}&=(-1.87,0.38)\,\rm{mas/yr}\\
(\mu_{\rm{W}},\mu_{\rm{N}})_{{\rm{SMC}}}&=(-1.08,-1.04)\,\rm{mas/yr}
\end{align}
\label{eq:BeslaPMresults}
\end{subequations}

The orbit around the Milky Way for the above set of parameters is shown in the left column of Figure \ref{fig:OrbitInfo}. The first panel describes the orbit of the Clouds around the Milky Way, showing that the Clouds were already within the virial radius of the Milky Way $3.0\;\rm{Gyr}$ ago. At this time, both galaxies were at a distance of $200\,\rm{kpc}$ from the Galactic Centre with a mutual separation of $30\,\rm{kpc}$. 

\begin{figure*}
\centering
\includegraphics{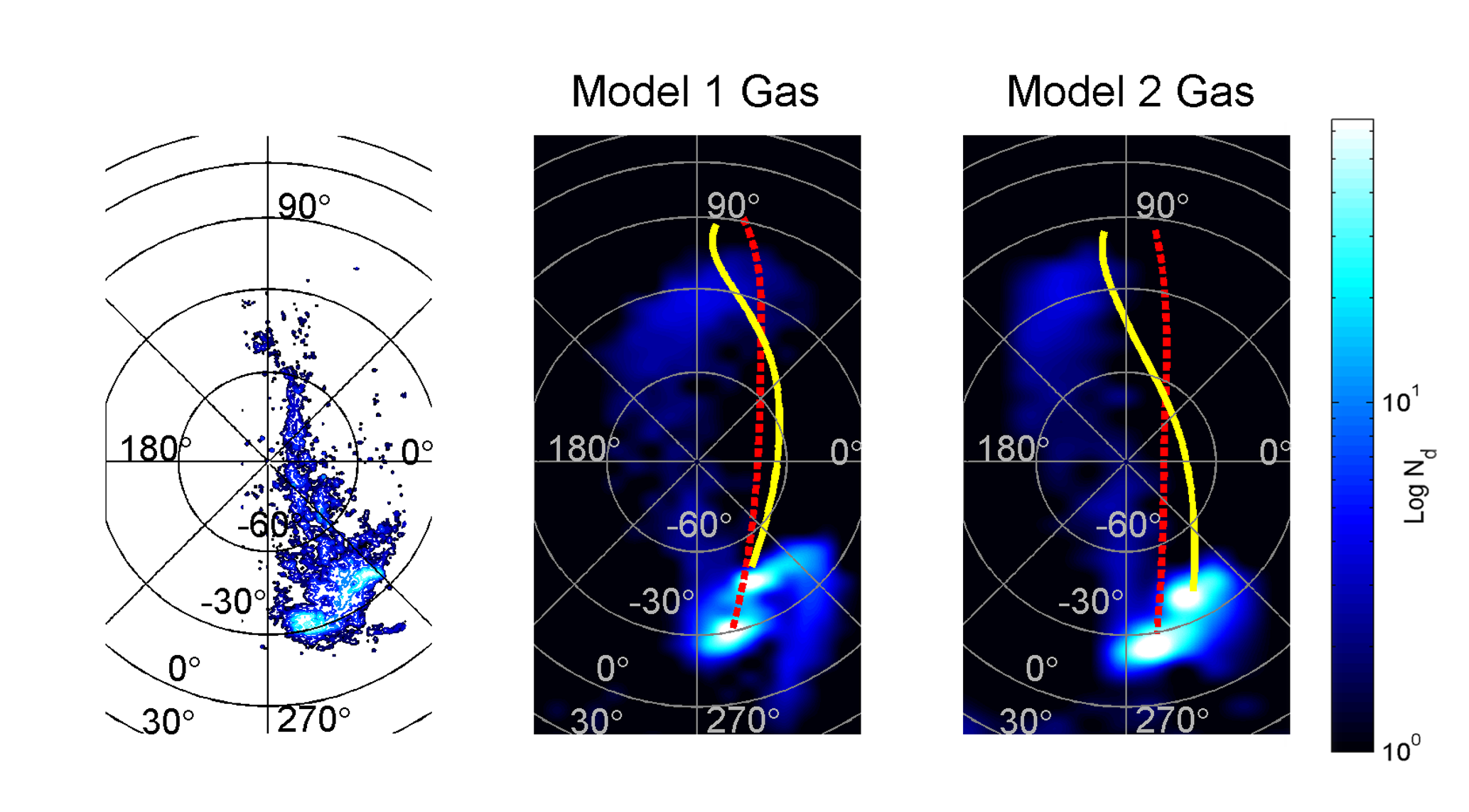}
\caption{ Full sky ZEA projection of the Magellanic System, centred on the South Galactic Pole. (From left to right) The observed HI column density of the Magellanic System as presented by \protect\cite{2003ApJ...586..170P}; final distribution of the gas particles in the simulated system are plotted for Model 1 (middle panel) and Model 2 (right panel). The dashed red and solid yellow lines are the projected orbit of LMC and SMC respectively in the last 1.5 Gyr, when the Stream starts to form.}
\label{fig:Zea}
\end{figure*}

\begin{figure*}
\centering
\includegraphics{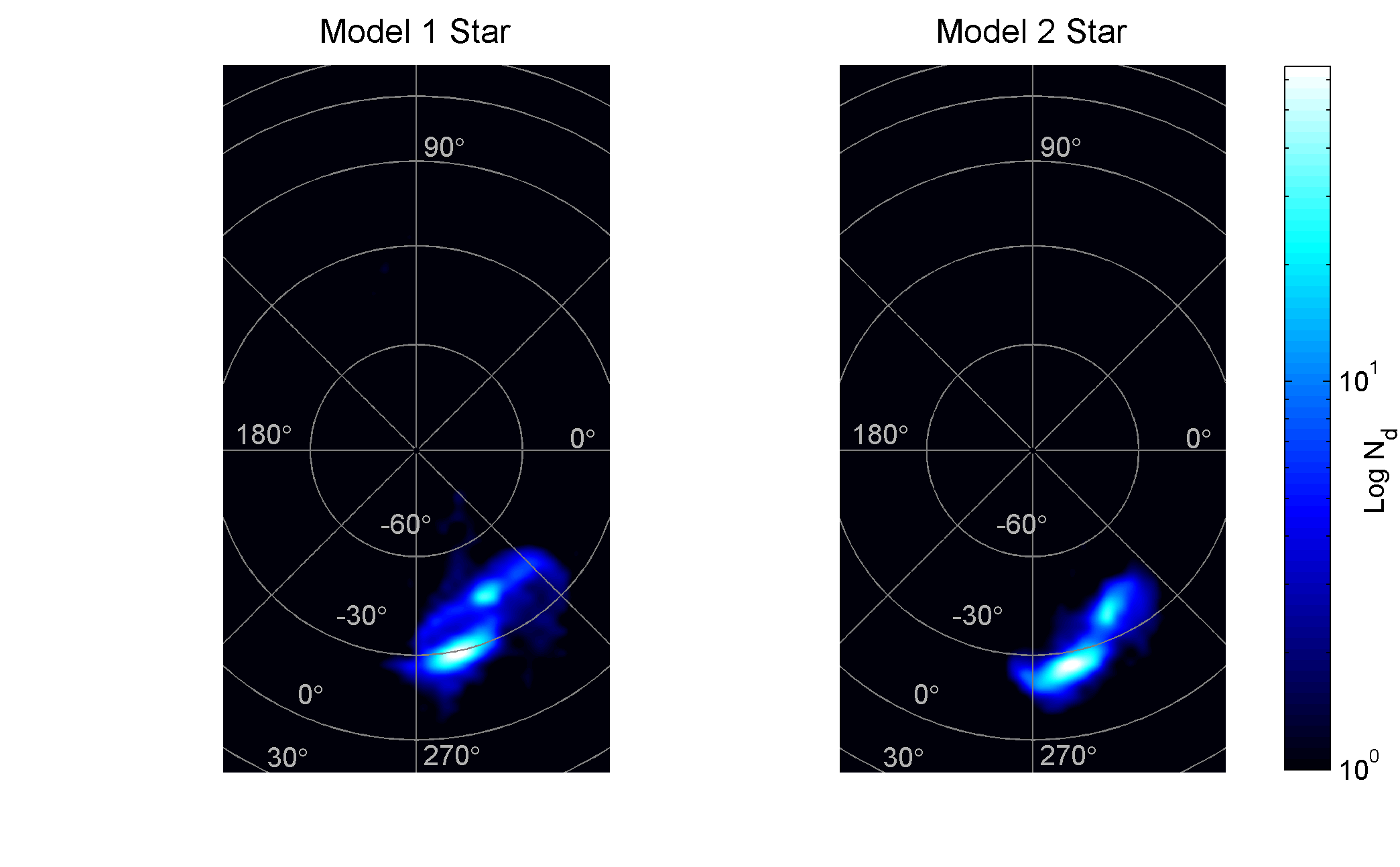}
\caption{Full sky ZEA projection of the star particle distribution in the Magellanic System  centred on the South Galactic Pole. As for Figure \ref{fig:Zea}, the final distribution of the star particles are plotted for Model 1 (left panel) and Model 2 (right panel). In both models, a bridge of stars connects the two galaxies, showing that both stars and gas have been tidally stripped during the last closer encounter between the Clouds.}
\label{fig:ZeaStar}
\end{figure*}

\begin{figure*}
\centering
\includegraphics{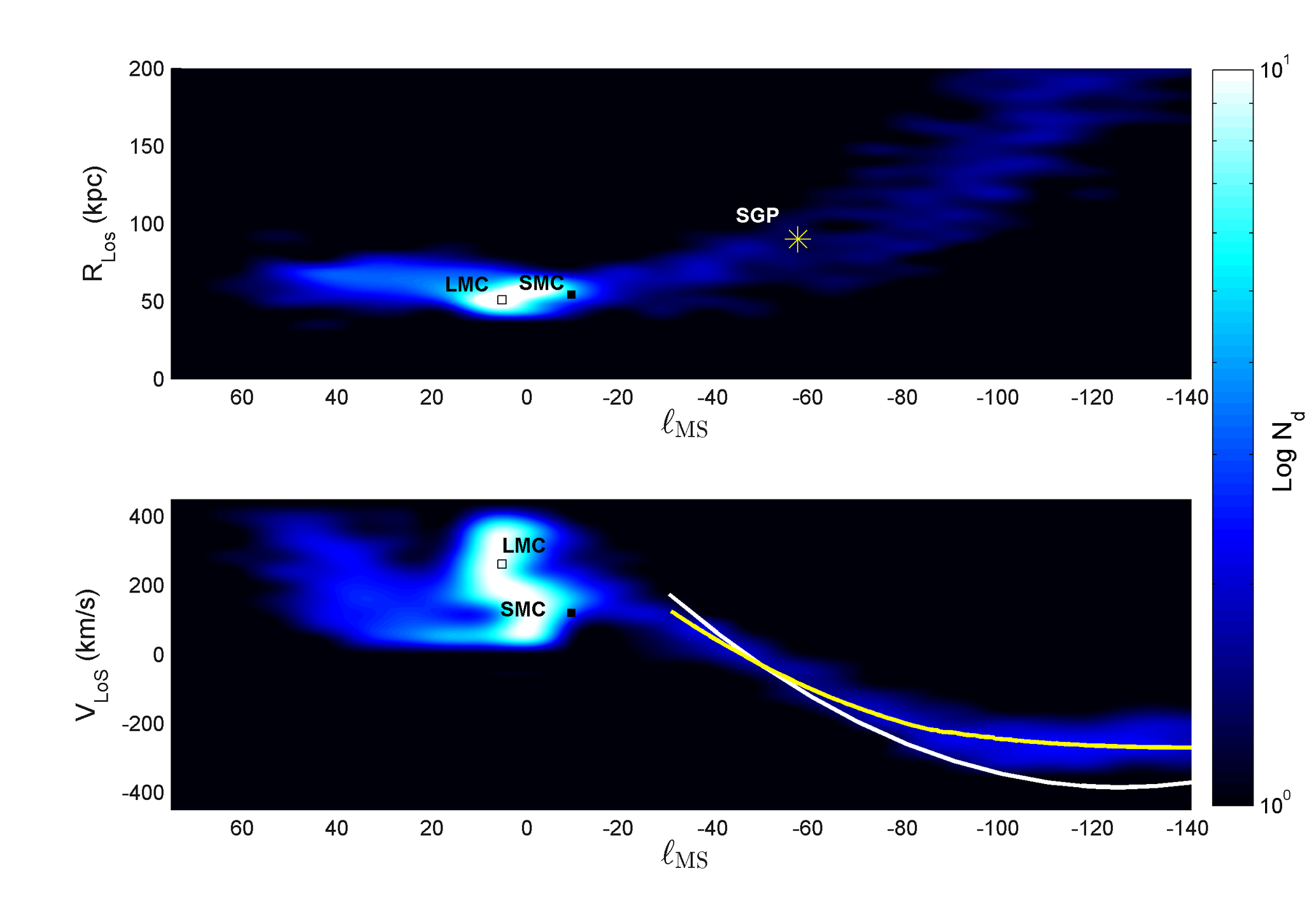}
\caption{ Line-of-sight distance (\emph{top panel}) and line-of-sight velocity (\emph{bottom panel}) for gas particles plotted as a function of the Magellanic Longitude, for Model 1. The \emph{white line} and \emph{yellow line} show the result of a polynomial fit applied on the data from \protect\cite{2010ApJ...723.1618N} (\emph{white line}) and on the simulated data (\emph{yellow line}). In the top panel, the \emph{yellow star} indicates the direction of the South Galactic Pole.}
\label{fig:BeslaVel}
\end{figure*}
The first panel on the second row in Figure \ref{fig:OrbitInfo} shows the distance between SMC and LMC as a function of time. SMC lies on a bound orbit around LMC, with a mean distance from the main Cloud always less than 50 kpc, and the mean mutual velocity less than 90 $\rm{km\,s}^{-1}$.  The first encounter between the Clouds occurs at $\rm{T}\approx-2.5\,\rm{Gyr}$, when SMC is $25\,\rm{kpc}$ away from the centre of LMC. This encounter is strong enough to change the morphology of the gas disc of SMC and form a temporary bridge of gas connecting the two Clouds. This structure lasts until the SMC starts to move away from LMC. Before the second encounter, SMC gas particles form arm-like structures. These lead to the formation of the Stream. The second encounter, $T=-0.38$ Gyr, is the strongest one, with the Clouds at a distance of $5\,\rm{kpc}$ from each other. 

The final configuration of the gas particles is shown in the middle panel of Figure \ref{fig:Zea}. As a result of the interaction between the Clouds, the final distribution of particles is such that the main components of the Magellanic System are reproduced.  When compared with the observed system (right panel in Figure \ref{fig:Zea}), the model well reproduce the position of the gas cloud at the position of the South Galactic Pole. However, the model present an over density at  $(\ell,b)\approx(110, -40)$ which is not present in the observed stream (left panel).

Figure \ref{fig:BeslaVel} shows the line-of-sight distance (top panel) and velocity (bottom panel) along the simulated Stream as a function of the Magellanic Longitude, $\ell_{\rm{MS}}$ \citep{2008ApJ...679..432N}. In both panels, the white star indicates the position of the South Galactic Pole (SGP). The line-of-sight distance increases along the Stream, having a minimum ($d=62\,\rm{kpc}$) at $\ell_{\rm{MS}}=-30^{\circ}$. The distance of the Stream at the position of the SGP is $78\,\rm{kpc}$. 

\begin{figure}
\centering
\includegraphics{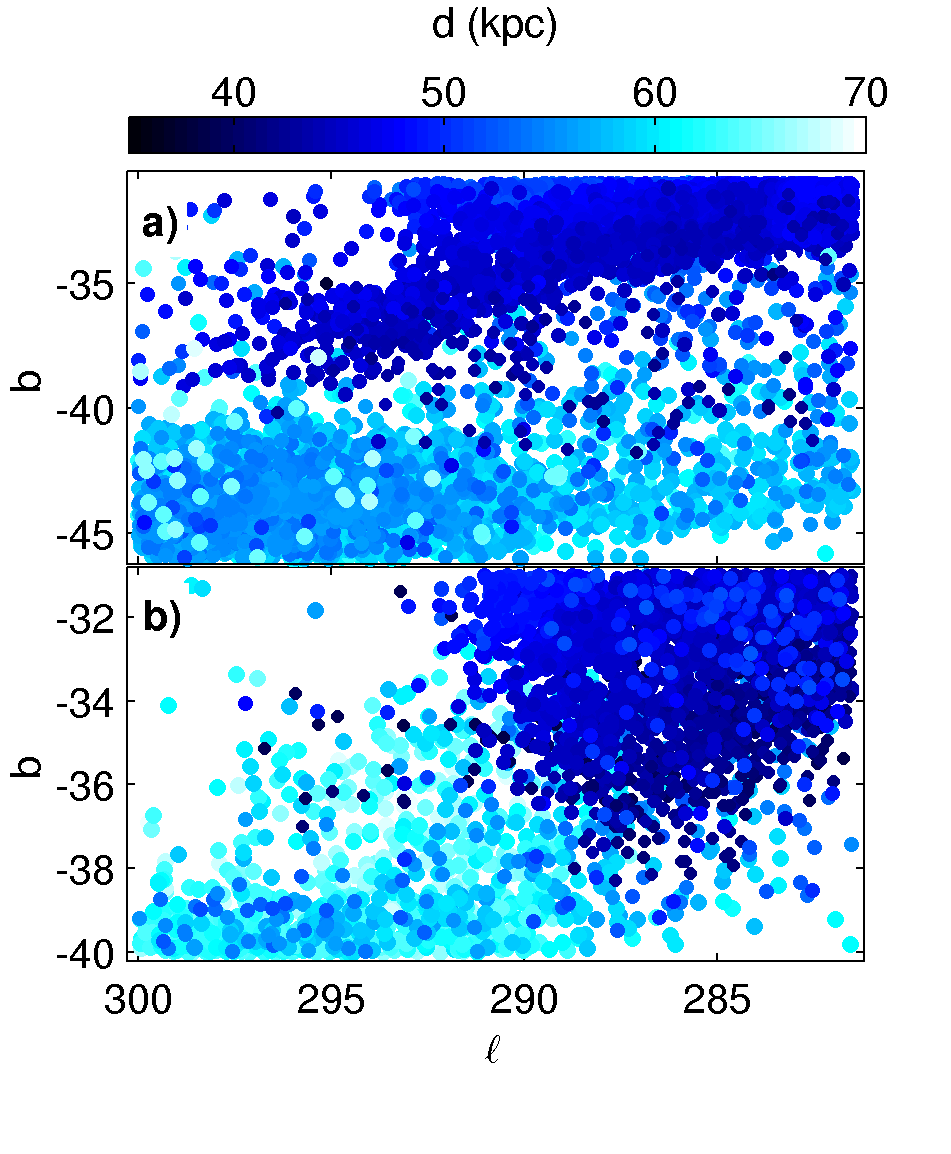}
\caption{ Stellar counter part of the Magellanic Bridge. This figure shows the distribution of the star particles in the intra clouds region, for Model 1 (panel a) and Model 2 (panel b). The colour map indicates the heliocentric distance of each particle.}
\label{fig:StellarBridge}
\end{figure}

\begin{figure*}
\centering
\includegraphics{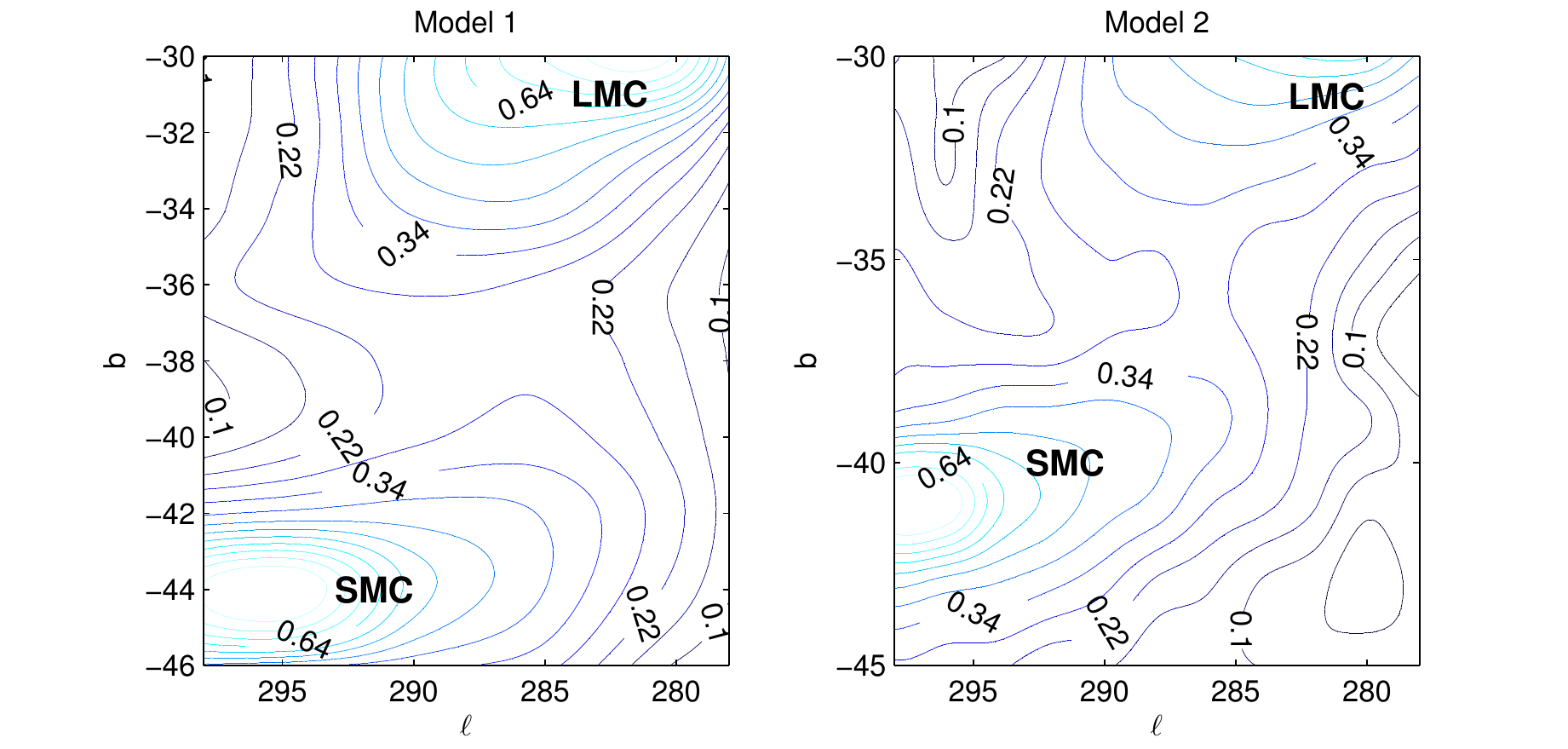}
\caption{Contour plot of intra cloud region  for Model 1 and Model 2. The contour lines refer to the percentage of the normalised density in the region between $281.5<\ell<300.0$ and $-47<b<-30$. Both models reproduce the location of the Magellanic Bridge.}
\label{fig:ModelBridge}
\end{figure*}

In addition to the Magellanic Stream, the selected model is able to reproduce the Magellanic Bridge as shown in the left panel of Figure \ref{fig:ModelBridge}: this feature is the result of the later encounter between the two Clouds. While the Stream is mainly formed by gas stripped from the SMC, both Clouds contributed to the formation of the Bridge. As the second encounter occurred at a distance of only 5 kpc between them, stars from both Clouds have been stripped and lied in this region. The bridge of star is clearly visible in the left panel of Figure \ref{fig:ZeaStar} in the intra clouds region (${282}^{\circ}\la{\ell}<290^{\circ}$ and ${-40^{\circ}}\la{b}<-32^{\circ}$). 
This seems to support the possibility of a presence of a stellar component of the Bridge. From the left panel in this figure, the SMC shows an extended tail in the North West direction, suggesting the presence of the star structure \citep{2014MNRAS.442.1680D}. Similar results are presented in \cite{2012ApJ...750...36D}, who suggested the presence of stars in the Bridge. 

To date, only a young population of stars have been confirmed in this region and they are believed to be formed in situ \citep{1990AJ.....99..191I}. Recently, \cite{2013A&A...551A..78B} suggested the presence of a older population of stars (with age between 400 Myr and 5 Gyr) in the Bridge. \cite{2013ApJ...779..145N} studied the SMC stellar periphery, showing the presence of stars in the eastern region of SMC extended in the direction of the LMC. They conclude that these stars are consistent with an intermediate/old stellar contour part of the HI Magellanic Bridge. 

In Figure \ref{fig:StellarBridge}, panel a) shows the presence of two distinct population of stars, with different heliocentric distance to indicate the contribution of the LMC ($d\sim50$ kpc) and SMC ($d\sim60$ kpc). The two populations extend from the two Clouds. Although the model present a distance bi-modality, this might not be consistent with the results presented by \cite{2013ApJ...779..145N}, as due to the different colour of LMC, it is unlikely that the stellar population in the SMC eastern ``stellar structure'' \citep{2013ApJ...779..145N}. However, as suggested by our simulations, LMC stars might also be observed in the region. 

%%%Figure6
\begin{figure*}
\centering
\includegraphics{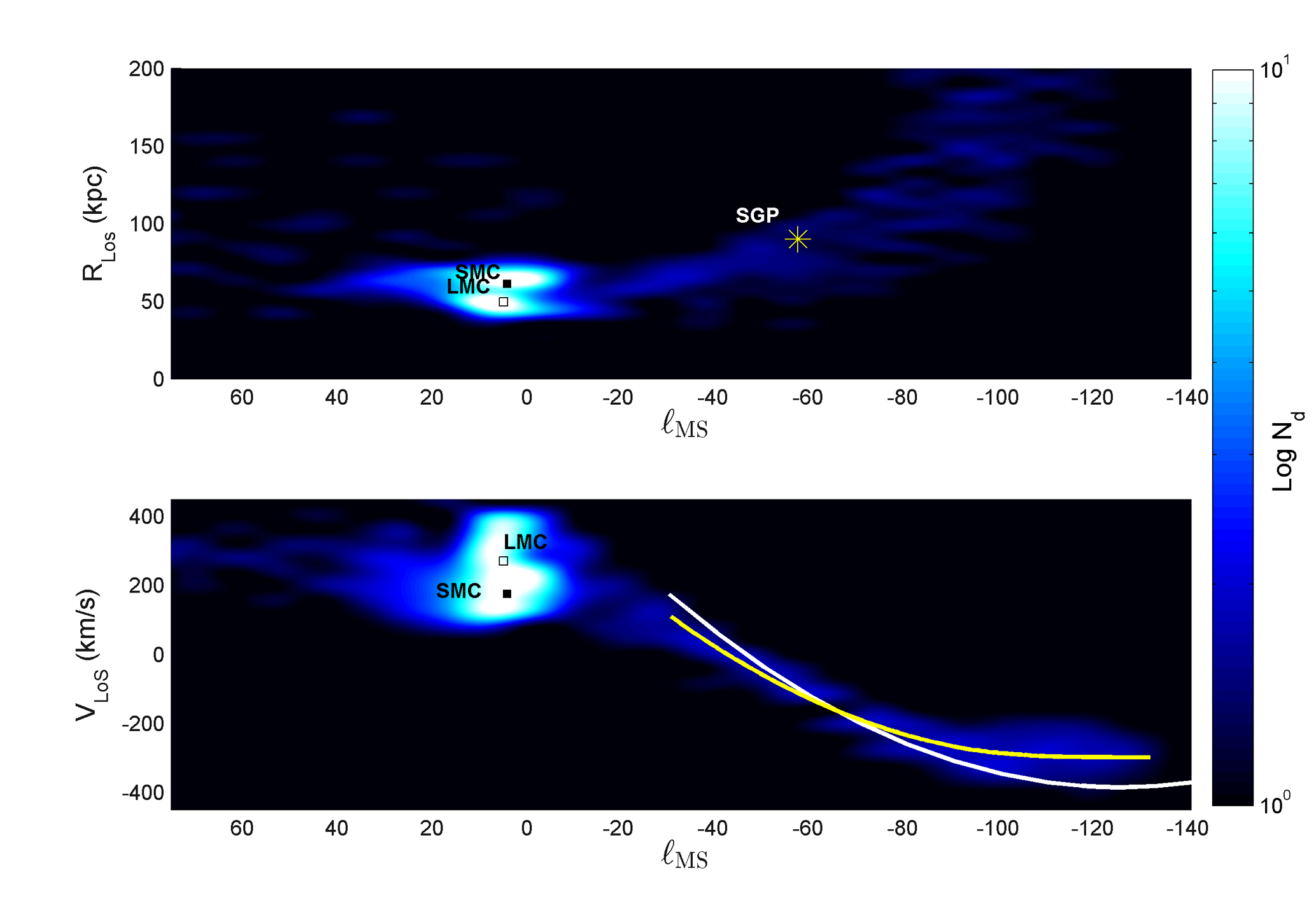}
\caption{ Model 2 results for the line-of-sight distance (\emph{top panel}) and velocity (\emph{bottom panel})for gas particles plotted as function of the Magellanic Longitude. As in the bottom panel in Figure \ref{fig:BeslaVel}, the white line in the bottom panel shows the fit on the data from \protect\cite{2010ApJ...723.1618N}, while the fit on the simulated data is plotted in yellow. The \emph{ yellow star} in the top panel shows the direction of the South Galactic Pole.}
\label{fig:Model2Vel}
\end{figure*}
%%%%%%%%%%%%%%%%%%%%%%%%%%%%%%%%%%%%%%%%%%%
The analysis of the observed line-of-sight velocity of the Stream as a  function of the Magellanic Longitude shows the presence of a velocity gradient between $-150^{\circ}\le{\ell_{\rm{MS}}}\le-30^{\circ}$, \citep{2003ApJ...586..170P,2010ApJ...723.1618N}. A clear gradient of the line-of-sight velocity is shown in the bottom panel of Figure \ref{fig:BeslaVel}. The white line shows the fit to the data from \cite{2010ApJ...723.1618N}; while the fit to the simulated data is plotted in yellow. The model reproduces the observed velocity range of the full system, with significant agreement with the observed fit.
%
% Fitness Vs Paramters
%
%
\begin{figure*}
\centering
\includegraphics{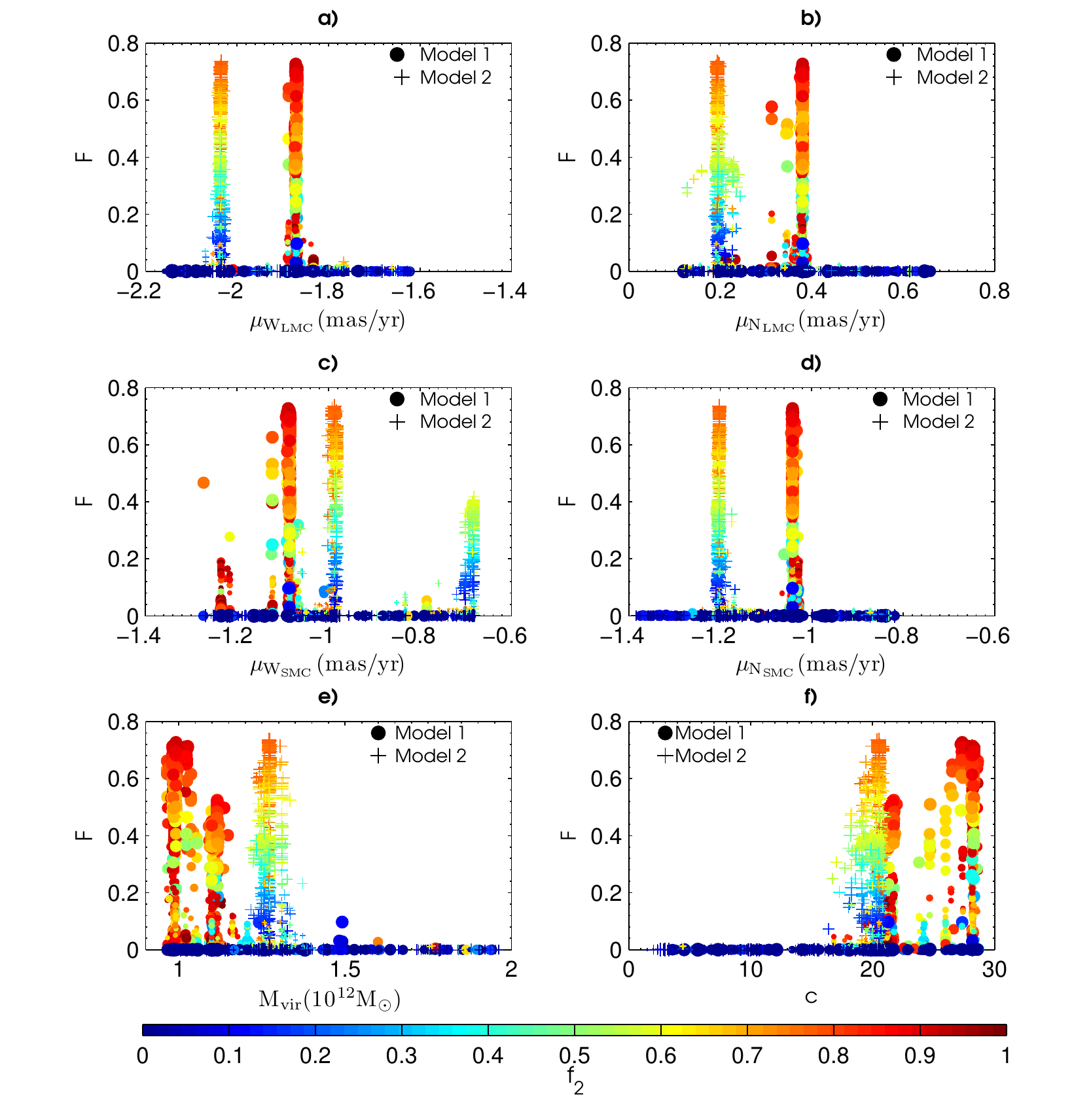}
\caption{ Distribution of the total fitness values as a function of each parameter, for Model 1 (\emph{filled circles}) and Model 2 (\emph{crosses}). In all panels, the size of the symbols is scaled with the values of the $f_1$ term; a greater size implies a good match in position and velocity of the simulated Clouds with the corresponding values from observations. The colour code shows the values of the $f_2$ component of the fitness function, so that blue points correspond to orbits which do not reproduce the requirements of the two encounters $\sim-2.5$ Gyr ago and $\sim-0.6$ Gyr ago. Panels a, b,c and d show the distribution of the fitness as a function of the west and north components of the proper motion for LMC and SMC. The dependences on the Milky Way halo parameters, virial mass and concentration are shown in panels e and f. }
\label{fig:GAParametervsFitness}
\end{figure*}
%%%
%
% ----------------SECOND MODEL--------------
%
\subsection{Model 2: A Better Orbital Model}
\label{sec:PrajSolution}
For this second model, the mass of the disc is $\rm{M_{disc}}=7.6\times10^{10}\,\rm{M_{\odot}}$ with a scale length of $\rm{r_{disc}}=6.5\,\rm{kpc}$ and a height equal to $0.26\;\rm{kpc}$, consistent with the recent estimation from 
\cite{2014Kafle} and the model described by \cite{2005ApJ...635..931B}.
As in the previous model, the mass and the concentration parameter for the dark matter halo are free parameters. For the best solution, these two parameters are $\rm{M_{vir}}=1.27\times10^{12}\;\rm{M_{\odot}}$ and $c= 20.5$, with a corresponding virial radius of $\rm{R_{vir}}=279\,\rm{kpc}$. The rotation curves are plotted in the left panel of Figure \ref{fig:RotCurve}.
The proper motions are 
\begin{subequations}
\begin{align}
(\mu_{\rm{W}},\mu_{\rm{N}})_{{\rm{LMC}}}&=(-2.03,0.19)\,\rm{mas/yr}\\
(\mu_{\rm{W}},\mu_{\rm{N}})_{{\rm{SMC}}}&=(-0.98,-1.20)\,\rm{mas/yr}
\end{align}
\label{eq:BeslaPMresults}
\end{subequations}
The plots in the second column of Figure \ref{fig:OrbitInfo} show the orbit of the Clouds in the last $3\,\rm{Gyr}$. This model presents several similarities to Model 1, although the mass model of the Galaxy is different. The Clouds are within the virial radius back to 3 Gyr. No close encounters with the Milky Way occurred during this time interval, except for the present day distance ($\rm{r_{LMC}}=49.1\,\rm{kpc}$ and $\rm{r_{SMC}}=59.1\,\rm{kpc}$). The SMC lies on an orbit of $0.62$ eccentricity around the Milky Way, while the LMC orbit has eccentricity of $0.64$. 
 As in the previous case, SMC is bound to LMC with a first encounter between LMC and SMC occurring at $T=-2.6\,\rm{Gyr}$ with a distance between them of $36\,\rm{kpc}$. Even with a greater separation than the previous model, this encounter is still strong enough to strip away gas from SMC, leading to the formation of a gas tail in the following $2\;\rm{Gyr}$. The last encounter at $T=-0.3\,\rm{Gyr}$ is more recent than the previous model, but still consistent with the reference values used in equation \ref{eq:f2}, with a mutual distance between the Clouds of $6\;\rm{kpc}$. As with Model 1, during this recent encounter between the LMC and SMC, stars and gas are tidally stripped from both Clouds to form the Magellanic Bridge, as shown in the right panel of Figure \ref{fig:ZeaStar}. Panel b in Figure \ref{fig:StellarBridge} shows the spatial distribution of the stars in the Bridge region. Despite being less distinct than Model 1, the  contribution of each Cloud is still observed as a difference in distance. The left panel in Figure \ref{fig:ModelBridge} shows a close look in the intra-clouds region, showing the gas distribution in the intra clouds region.

The final configuration of the gas  particles in the all sky  projection is provided in the right panel of Figure \ref{fig:Zea}. Although there is clear evidence of an extended tail in the position of the Magellanic Stream, for this model there is not a well formed Leading Arm. Figure \ref{fig:Model2Vel} shows the line-of-sight distance along the stream (top panel) and the gradient of the line-of-sight can be seen in the the bottom panel of the same Figure. As for Model 1, the distance along the Stream increases with $\ell_{\rm{MS}}$, having a minimum value of $63\;\rm{kpc}$ at $\ell_{\rm{MS}}=-30^{\circ}$ and a distance of $76\;\rm{kpc}$ at the SGP.

%
%Table Mass 
%
%
%
\begin{table*}
\centering
\begin{tabular}{ccccccc}
\hline
Galaxy & & Initial ($10^9$ M$_\odot$) &Model 1($10^9$ M$_\odot$) & Model 2 ($10^9$ M$_\odot$)& Observed ($10^9$ M$_\odot$)&Reference\\
\hline
LMC& Star ($<9$)& 2.16&2.14&2.13&2.7&\cite{2002AJ....124.2639V}\\
&Gas($<5$)&0.83&0.82&0.82&0.44&{\cite{2005A&A...432...45B}}\\
&Total($<9$)&19.1&19.0&19.0&17&\cite{2014ApJ...781..121V}\\
\hline
\hline
SMC& Star ($<3$)&0.24&0.21&0.17&0.31&\cite{2004ApJ...604..176S}\\
&Gas($<3$)&0.58&0.60&0.60&0.42&\cite{2004ApJ...604..176S}\\
&Total($<3$)&2.84&2.24&2.21&2.7&\cite{2006AJ....131.2514H}\\
\hline
\hline
Stream&Gas &-&0.01& 0.01&0.02&\cite{2010ApJ...723.1618N}\\
Bridge&Gas &-&0.5&0.3&0.25-5.0 &{\cite{2005A&A...432...45B}}, \citet{2013ApJ...771..132B}\\
&Star &-&0.09&0.09&- &-\\
\hline
\end{tabular}
\caption{Initial and final mass distribution for Model 1 and Model 2, compared with expected results from observations. The Stream is defined as particles with $\ell_{\rm{MS}}<-30^{\circ}$. The Bridge mass is calculated considering all the gas particles (and stars) in the region $281.5<\ell_{\rm{MS}}<300.0$ and $-44<\ell_{\rm{MS}}<-31$. }.
\label{tab:MassClouds}
\end{table*}
\subsection{Mass distribution}

Table \ref{tab:MassClouds} summarises the initial and final mass contained within the characteristic radii for the LMC and SMC. The quoted values from literature are also indicated in the last two columns.

Both Model 1 and Model 2 are generally in good agreement with the observations. However, the final gas mass estimated for both Clouds is greater than the expected values. For the case of LMC in particular, the gas content within 5 kpc is double the quoted values. For the initial conditions, we deliberately included much more gas than expected in order to account for the gas loss due to the interaction with SMC and the Milky Way. However, during the interaction between SMC and the Galaxy in both models, the gas content in the inner region of the LMC disc remains untouched. In addition, it is important to note that no star formation prescription has been added to the simulations. Introducing star formation in the models presented in this work it would reduce the inner gas reservoir of the Clouds.

Conversely, the mass of the Stream is underestimated when compared with the value presented by \cite{2010ApJ...723.1618N}. This could be due to the lack of ram pressure in the models, which can influence the final mass distribution of this component. Moreover, the inclination of the SMC's disc with respect to LMC can play a significant role in the formation of the Magellanic Stream and can also influence the final mass content in this structure \citep{2012MNRAS.421.2109B}.

As consequence of the last encounter between the two Clouds, a bridge of gas is seen connecting the two Clouds. Following \cite{2013ApJ...771..132B}, we defined the intra-clouds regions to be ${281.5^{\circ}}\la{\ell}<~300.0{^{\circ}}$ and ${{-44}^{\circ}}\la{b}<-32{^{\circ}}$ and we calculated the mass of the Bridge by considering gas particles which lie in this region. This leads to a Bridge mass of $5\times{10^{8}}\,{\rm{M_{\odot}}}$ for Model 1 and $3\times{10^{8}}\,{\rm{M_{\odot}}}$. Using the Parkers HI Survey, \cite{2005A&A...432...45B} estimated the mass within the Bridge to be $2.5\times{10^{8}}\,{\rm{M_{\odot}}}$. \cite{2013ApJ...771..132B} used the Wisconsin H$\alpha$ Mapper to resolve the warm ionised gas component in the Magellanic Bridge. They showed that the mass of the ionised material varies between $0.7-1.7\times{10^{8}}\,{\rm{M_{\odot}}}$. They also proposed a estimation of the neutral mass in the same regions $3.3\times{10^{8}}\,{\rm{M_{\odot}}}$. In the last column of Table \ref{tab:MassClouds}, we used the results of \cite{2005A&A...432...45B} as minimum mass in the region and the total gas mass (neutral + ionised) given by \cite{2013ApJ...771..132B} as upper limit.

One limitation of this analysis is that the N-body simulations do not allow us to explore different mass models for the Clouds. The results presented here are therefore model-dependent, as different initial masses of the Clouds can lead to different orbital scenarios, with different choices for the best parameters of the Milky Way.
Although \cite{2014ApJ...781..121V} show that the dynamical mass of LMC within 9 kpc is 1.7$\times10^{10}$ M$_\odot$, they suggest that this value can underestimate the total mass of LMC and that a different virial mass is possible if the concentration of the halo is adjusted to reproduce the rotation curve. \cite{2013ApJ...764..161K} explore orbit configuration for different values of the LMC mass, increasing it up to 25$\times10^{10}$M$_\odot$. A further investigation could be to repeat the GA analysis, using different combinations for the masses of the Clouds, in order to have a more complete picture of their interaction with the Milky Way.

\section{Fitness dependence on parameters}
\label{sec:Fparameters}

In this section, we describe the influence of each parameter on the fitness function of the two models analysed. Figure \ref{fig:GAParametervsFitness} shows the dependence of total fitness ($F=f_1*f_2*f_3$) on the GA parameters for Model 1 (filled circles) and Model 2 (crosses). Each panel in this Figure corresponds to one of the six parameters as indicated on the x-axis. The blue-to-red colour map indicates the values of $f_2$ (two encounters between the Clouds at the given time, see \S\ref{sec:MF}). Blue points correspond to orbits in which no encounters or only one encounter between the Clouds occurred in the last 3 Gyr; while red points correspond to orbits with at least two encounters, one at T$\sim{-2.5}$ Gyr and one at T$\sim{-0.4}$.

The first four panels in Figure \ref{fig:GAParametervsFitness} show the distribution of the peaks in the total fitness ( over the studied range $\pm1\sigma$) of the proper motion in the northern and western directions of both Clouds. The proper motions are considered free to span in the error range given by \cite{2010AJ....140.1934V}. 

Panels a and b show the fitness as a function of the west and north proper motion components for LMC. In both models, it is possible to distinguish two clear peaks of the fitness function, corresponding to the two Milky Way models. Panel a shows that the selected individuals lie in the interval $ \mu_{\rm{W_{LMC}}}\in[-2.16, -1.89]\,\rm{mas\,yr^{-1}}$, which corresponds to the $-1\,\sigma$ in the \cite{2010AJ....140.1934V} catalogue. On the other hand, all the rejected solutions fall in the region $[-1.8, -1.6]\,\rm{mas\,yr^{-1}}$. Similarly, panel b shows that all the solutions belong to the region $\mu_{\rm{W_{LMC}}}\in[0.2, 0.4]\,\rm{mas\,yr^{-1}}$.

The dependence of the fitness function on the SMC proper motion is shown in panels c (west component) and d (north component). As shown in panel c, the fitness function peaks in the regions $\mu_{\rm{W_{SMC}}}\in[-1.1, -0.98]\,\rm{mas\,yr^{-1}}$. However, Model 2 (crosses) identifies a local maximum around -0.78 mass$\,$yr$^{-1}$. This corresponds to a total fitness value of 0.4, around the 21st generations which has been later rejected. As for LMC, panel ) shows that the best solutions for the north components of SMC fall in $-1\sigma$ region (range [-1.2 -1.0] mass$\,$yr$^{-1}$). 

From the results in panels a to e, we can constraint the relative proper motions between the Clouds. Both models lead to a value of the proper motion in the north direction between the range $\mu_{\rm{N_{LMC-SMC}}}=1.40\pm0.02\,\rm{mas\,yr^{-1}}$, consistent with the quoted value $\mu_{\rm{N_{LMC-SMC}}}=1.49\pm0.15\,\rm{mas\,yr^{-1}}$ in \cite{2010AJ....140.1934V}. In contrast, in the west direction for Model 1 the values span between $\mu_{\rm{W_{LMC-SMC}}}=[-0.88,-0.77]\,\rm{mas\,yr^{-1}}$; and for Model 2 $\mu_{\rm{W_{LMC-SMC}}}=[-1.06,-1.03]\,\rm{mas\,yr^{-1}}$, in agreement with the quoted value 0.91$\pm0.16\,\rm{mas\,yr^{-1}}$ \citep{2010AJ....140.1934V}.

The last two panels e and f in Figure \ref{fig:GAParametervsFitness} provide the distribution of the fitness as a function of the virial mass and the halo concentration parameter, respectively. Despite the difference in the disc and bulge parameters adopted for the two models, in both cases the GA selects virial mass less than $1.5\times10^{12}$ M$_\odot$ (see panel e). The peaks in the virial mass correspond to high values for the concentration, as shown in panel f. 

The differences between Model 1 and Model 2 that can be deduced from panel f are interesting. For Model 2 (crosses), a single peak is clearly observed at the best values selected by the GA ($\rm{M_{vir}}=1.27\times10^{12}$ M$_{\odot}$ and c$=20$). The distribution for Model 1 instead shows two peaks, corresponding to a total fitness of 0.73 and 0.53. These two peaks can be observed in both virial mass and concentration. The highest peak corresponds to the best individual in the last generations, while the second corresponds to a high ranked solution with a virial mass around $1.1\times10^{12}$ M$_{\odot}$ and concentration around 21. This double peak feature is not observed in Model 2, suggesting that the selection of the parameters for the disc and bulge potential play an important role in shaping the potential of the Milky Way, even if the component which strongly drives the orbit of the Clouds is the dark matter halo.

It is interesting that both models support scenarios where the Clouds are orbiting around a less massive, but more concentrated Milky Way. In section
\ref{sec:MWresults} we discuss about the implications of these findings.

\subsection{The Effect of the Dynamical Friction}
\label{sec:DF}

So far the dynamical friction between the two Clouds and the Milky Way has not been included in the genetic algorithm. However, the orbit of the Clouds might be strongly influenced by the interaction  with the dark matter halo of the Milky Way. 
In this section we investigate the influence of the dynamical friction on the selected orbits discussed above. 

We modified the equation of motion in Gadget2 in order to introduce the dynamical friction term, assumed to act on each particles. The dynamical friction is give by the Chandrasekhar's formula 
\begin{equation}
\frac{d\mathbf{v}}{dt}=-\frac{4\pi\,G^2\,{m{_{i}}{^2}}\,\ln{\Lambda}\,{\rho(r)}}{v^2}\left[{\rm{erf}}(X)-\frac{2X}{\sqrt{\pi}}\,\exp(-X^2)\right]\frac{\mathbf{v}}{v},
\label{eq:DF}
\end{equation}
where $v$ is the orbital velocity of each particles and $X=v/\sqrt2\,\sigma$, with $\sigma$ being the one-dimensional velocity dispersion of the adopted NFW dark matter halo. For a given position, we calculate the value of $\sigma$ using the numerical approximation described by \cite{2003ApJ...598...49Z}. Following \cite{2003ApJ...582..196H}, the value of the Coulomb logarithm is defined to vary as a function of the distance, given by $\Lambda=r/1.6\epsilon$, where $\epsilon$ is the softening length if the Clouds were modelled using a Plummer profile. In particular, we used $\epsilon=5.0\,\rm{kpc}$ for LMC and $\epsilon=3.0\,\rm{kpc}$. \cite{2003ApJ...582..196H} show that this prescription reduce the discrepancy between semi-analytic approximation and N-Body simulation.

The dashed lines in Figure \ref{fig:OrbitInfo} show the influence of the dynamical friction on the best orbital models. In general, the initial distance of the two Clouds from the Milky Way are greater than the case without dynamical friction. However, both Clouds were still within the Milky Way virial radius, shown as a dashed grey line. 

The effects due to the dynamical friction are stronger on the LMC, but as the orbit of the latter changes, the SMC orbit changes as well. The central panels in Figure \ref{fig:OrbitInfo} show the comparison between the evolution of the SMC around LMC in both models as a function of time, for both cases, with (solid line) and without (dashed line) dynamical friction. In spite of the difference in the relative orbit, the two Clouds remained a bound pair for the last 3 Gyr and as for the case of no dynamical friction, two encounters between the Clouds occurred in this time. Therefore, the selected orbits are still good solutions for the genetic algorithm. In addition, no significant changes in the final configuration of the Magellanic System are observed when the dynamical friction is added. Both models in fact are still able to reproduce the main features of this system. This is not surprising because as shown in the previous sections, having the Clouds bound to each other and requiring them to interact for at least 3 Gyr are necessary conditions to lead the formation of the main features of the Magellanic System. These two conditions are still satisfied by the orbit obtained with the dynamical friction.

\subsection{The Orbit of SMC around LMC}

%
% Fitness Vs Ecc
%
\begin{figure}
\centering
\includegraphics{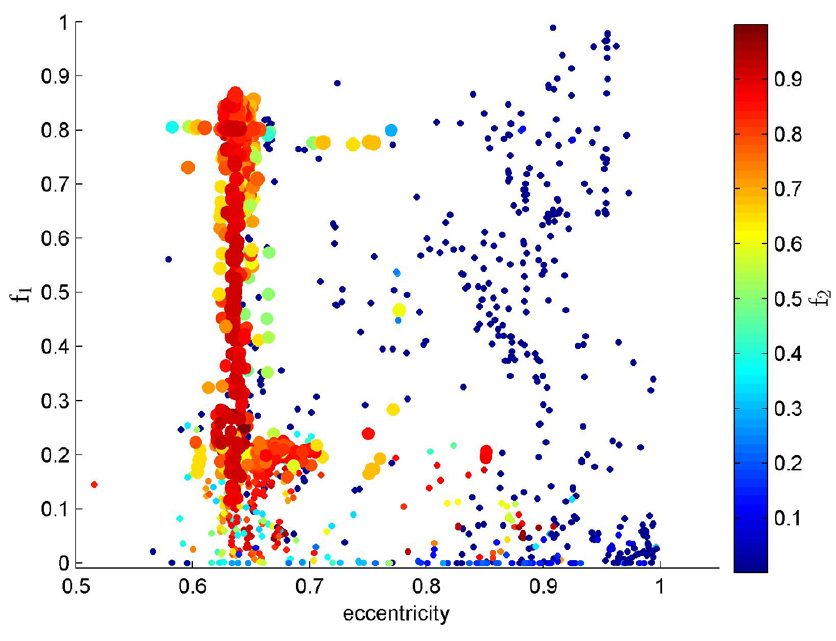}
\caption{ The dependence of the fitness function on the orbital eccentricity of the SMC around LMC. The size of the points is scaled according the values of the total fitness function (equation \ref{eq:TotFit}), while the colour map refers to the value of $f_2$. The final positions and velocities of both simulated Clouds is reproduced only for eccentricity between 0.6-0.7, as indicated by the values of $f_1$.}
\label{fig:F2vsEcc}
\end{figure}
%----------------------------------------
%----------------------------------------
%----------------------------------------
The genetic algorithm does not constrain the Clouds to be bound to each other or to the Milky Way, in order to avoid imposing any strong assumption on the history of these galaxies. The selection of the best orbit is based only on the results of the N-body simulations.

As discussed in \cite{2012MNRAS.421.2109B}, the SMC needs to orbit around the LMC with eccentricity of approximately $0.7$ to avoid extreme cases of fly-by or quick orbital decay with a subsequent merger.  In both cases, the final positions and velocities of SMC do not match the expected values. The genetic algorithm  selected an orbit by comparing the final position and velocities of both Clouds with their present day values. Hence, the fly-by/merger cases are automatically avoided, as they lead to low values of the $f_1$ term (see equation \ref{eq:f1}). This implies a low values for the fitness function. In this section, we investigate if there is any dependence on the eccentricity in the fitness function.

In Figure \ref{fig:F2vsEcc}, the $f_1$ values of each individual is plotted against the orbital eccentricity. The $f_1$ term indicates how well the final positions and velocities of the Clouds match the expected values for the corresponding orbit eccentricity. The colour (\emph{blue to red}) of each point indicates the values of the term $f_2$ (\emph{low to high}). This term contains information on the number of encounters between the Clouds and the time when they occurred. Therefore, blue points on this plot either correspond to an orbital scenario where the Clouds did not interact with each other at any time or only one encounter occurred. The size of each point is scaled according to the value of the total fitness, $F$. 

As shown in Figure \ref{fig:F2vsEcc}, all the solutions of the genetic algorithm peak around an eccentricity between $0.60-0.70$. Those orbits are able to simultaneously satisfy the condition on the encounters between the Clouds (\emph{red points}) and the present day position and velocity ($f_1\sim0.7$), corresponding to high values of the total fitness function.
 
The results for orbits with eccentricity greater than 0.8 are interesting. Those orbits can be separated in two different regions in Figure \ref{fig:F2vsEcc}:
\begin{itemize}
\item[-] \emph{High $f_1$- Low $f_2$ values}.
 These orbits are able to reproduce the present day position and velocity ($f_1\sim1$), but there are no encounters between the Clouds in the first $2\,\rm{Gyr}$, as indicated by the majority of blues points, (low value of $f_2$).  If no interactions occur between the Clouds, neither galaxy is tidally disrupted ( there is no formation of a particle stream); and therefore, their orbits do not deviate from those of point mass, resulting in an almost perfect match with their positions in the sky (a high value of $f_1$). Due to the low value of the total fitness the corresponding parameters are considered weak individuals by the genetic algorithm and do not survive for the next generation.
 \item[-] \emph{Low $f_1$-High $f_2$ values}.
 There are solutions corresponding to high eccentricity orbits in which SMC has multiple encounters with LMC. These orbits decay too quickly, leading to a full merger between the Clouds. (A similar fate is observed for systems with eccentricities lower than $0.5$.) 
\end{itemize}
The pronounced peak around 0.6-0.7 eccentricity suggests that such SMC orbits are necessary to ensure the survival of this Cloud \citep{2012MNRAS.421.2109B} and to have two encounters between the Clouds at $\rm{T}\sim-2.5\,\rm{Gyr}$ and $\rm{T}\sim-0.4\,\rm{Gyr}$ \citep{2009AJ....138.1243H}.

\section{Fundamental Parameters}
\begin{figure*}
\centering
\includegraphics{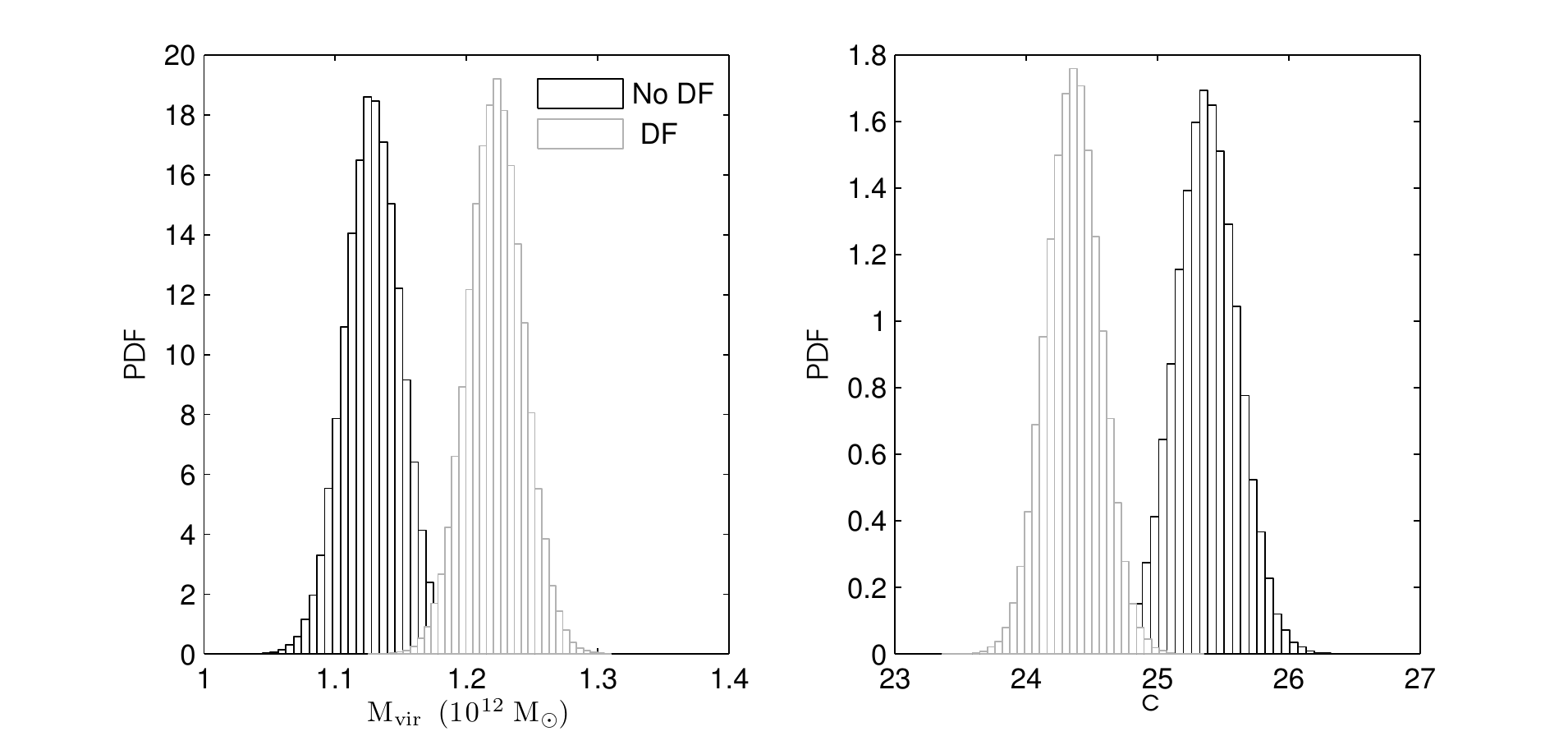}
\caption{From \emph{left to right}: The marginalised distribution of the virial mass (${\rm{M_{vir}}}$) of the Milky Way and its variance ($\sigma_{\rm{M_{vir}}}$) and the marginalised distribution for the concentration parameter, $\rm{c}$.}
\label{fig:Histo}
\end{figure*}
\subsection{The Virial Mass of the Halo}
\label{sec:MWresults}
The parameters related to the Milky Way halo potential are crucial for determining the orbital history of the Magellanic Clouds. As the circular velocity depends strongly on the contribution of the dark matter component, the halo parameters lead to different values of the circular velocity, which can influence the orbits of the Clouds \citep{2009MNRAS.392L..21S,2010ApJ...725..369R}.

The major source of uncertainty is the virial mass of the Milky Way. Analysis on the kinematics of the stellar halo of our Galaxy constrains this parameter to the range $0.9-1.5\times10^{12}\rm{M_{\odot}}$ \citep{2008ApJ...684.1143X,2012ApJ...761...98K}. However, dynamical models for the Magellanic Clouds favour  a more massive halo in order to justify the formation of the Magellanic System. \cite{2012ApJ...750...36D} show that a minimum values of $\geq{1.3\times10^{12}\,\rm{M_{\odot}}}$ for the virial mass is necessary to form the Magellanic Stream. 

Different models for the dark matter component influence the orbit of the Clouds around the Milky Way, since they change the gravitational field and  circular velocity at the solar distance. \cite{2013ApJ...764..161K} show that bound configurations between the LMC and SMC are more likely for high values of the LMC mass, but there is also a dependence on the virial mass of the Milky Way. The required LMC mass to keep the SMC bound increases as the Milky Way mass increases. As a consequence, the orbital eccentricity of the LMC changes, making a first infall scenario more likely.

In both models presented, the Clouds have formed a bound pair for a time interval of at least $3\,\rm{Gyr}$, even for low mass model of the Clouds.  We show that the bound configuration of the Clouds is crucial for the formation of the Stream. Since neither models have an encounter with the Milky Way, the only encounters between the SMC and LMC are strong enough to strip material from SMC, leading to the formation of the Stream \citep{2012MNRAS.421.2109B}, even for a low value of the virial mass of the Milky Way.

In order to characterise this finding, a set of 100 proper motions has been drawn from the error distribution of the \cite{2010AJ....140.1934V} catalogue. For each of these values, a genetic algorithm ran considering the two Clouds being Plummer spheres with radius 5 kpc and 3 kpc for the LMC and SMC respectively. Since all the good orbits found using the N-body simulations concentrate around an eccentricity of the SMC orbit around LMC between $0.6-0.7$, the $f_1$ term in equation \ref{eq:TotFit}, has been modified to be
\begin{equation}
 f_1=\frac{1}{1+\left(\frac{e_i-0.65}{0.05}\right)^2}
 \label{eq:f1}
\end{equation} 
while the condition on the encounters ($f_2$) and ($f_3$) are kept as described in \S\ref{sec:MF}. In each genetic algorithm run, 150 phenotypes evolve for 150 generations. The same analysis has been repeated using the disc and bulge parameters for Model 1 and Model 2.

Since in this analysis a point mass approximation is used, the equation of motion has been modified to explicitly account for the dynamical friction between the Clouds. Following \cite{2005MNRAS.356..680B}, this extra term is defined by \citep{2008gady.book.....B}

\begin{equation}
F_{\rm{DF}}=-0.428 \ln{\Lambda}\frac{\rm{G\,M_{\rm{SMC}}^{2}}}{r_{\rm{LS}}^2}\frac{\mathbf{v}_{\rm{LS}}}{v_{\rm{LS}}},
\label{eq:DFCL}
\end{equation}
where $r_{\rm{LS}}$ ($v_{\rm{LS}}$) is the distance (velocity) of SMC with respect LMC and $\ln{\Lambda}$ is the Coulomb logarithm chosen to be 0.2.

Once a distribution of the best individuals is established, we use an MCMC estimator to find the most likely values for the virial mass, the concentration of the Milky Way dark matter halo and their respective standard deviation. The prior for these two parameters is given by the parameter range described in Table \ref{tab:GAparam}.
The final distribution for Model 1 (virial mass and concentration) is shown in Figure \ref{fig:Histo}. The most likely value for the virial mass for Model 1 (Model 2) is $\rm{M_{vir}}=1.1\pm{0.2}\times10^{12}\,\rm{M_{\odot}}$ ($\rm{M_{vir}}=1.2\pm{0.2}\times10^{12}\,\rm{M_{\odot}}$), showing that solutions with a lighter Milky Way halo are preferred. However, this is possible only with higher concentration values. Indeed, the most likely value is $\rm{c}=25\pm2.0$ ($\rm{c}=20\pm2.0$).

Introducing a dynamical friction term between the Clouds and Galaxy in the equation of motion might influence the orbit of the Clouds and the time they spend in the Milky Way halo. In section \S\ref{sec:DF}, we have shown how the dynamical friction influence the orbit of the best solution selected by the genetic algorithm. There is however, the possibility that the dynamical friction might also influence the selection of the parameters related to the Milky Way halo. 

With the aim of understanding how this term might affect the outcome of the genetic algorithm, we repeated the analysis, but including the dynamical friction due to the Milky Way halo, using the same approach discussed in section \S\ref{sec:DF}. The grey histogram in Figure \ref{fig:Histo} shows the distribution of the virial mass (right panel) and concentration (left panel) for Model 1 when equation \ref{eq:DF} is included in the equation of motion. With the dynamical friction, the genetic algorithm tends to prefer a slightly higher mass ($\rm{M_{vir}}_{\rm{DF}}=1.2\pm{0.2}\times10^{12}\,\rm{M_{\odot}}$) and a slightly smaller values for the concentration parameters ($\rm{c}=24\pm2.0$).  Despite the differences, these results are still consistent with the quoted values for Model 1 and Model 2, confirming the early statement that the lighter and more concentrated halo are favourite solutions.

Results from numerical simulations show that the mass-concentration relation predicts a mean value for the concentration parameter of $10$, when the virial mass is around $10^{12}\rm{M_{\odot}}$ \citep{2008MNRAS.391.1940M}. Both results found by the genetic algorithm appear to overestimate this parameter. 
However, the relation between virial mass and concentration is based on dark matter simulations, while the presence of baryons can adiabatically contract the dark matter halo, leading to more concentrated haloes \citep{1998MNRAS.295..319M,2004ApJ...616...16G,2013ApJ...773L..32R}. In addition, the studies of the potential of the Milky Way based on the kinematic of its stellar halo all converge to the conclusion that a virial mass of the dark matter halo $\le10^{12}\rm{M_{\odot}}$ and a high concentration parameter are favoured over the less concentrated-more massive halo model \citep{2005MNRAS.364..433B,2012MNRAS.424L..44D, 2014Kafle}. 

The dependence of the genetic algorithm results on the particular choice of the disc and bulge mass of the Milky Way is not surprising. As shown in panel f of Figure \ref{fig:GAParametervsFitness}, the fitness as a function of the concentration peaks around 27 for Model 1 and 20 for Model 2. The two models differ only in the parameters of the disc and bulge, with Model 1 having a less massive disc and bulge with respect to Model 2. As a consequence, the halo concentration is greater when a less massive disc is considered. This suggests that there is an anti-correlation between the disc parameter and the dark matter halo concentration.  

\subsection{The Distance to the Magellanic Stream}
%Figure
 \begin{figure}
\centering
\includegraphics{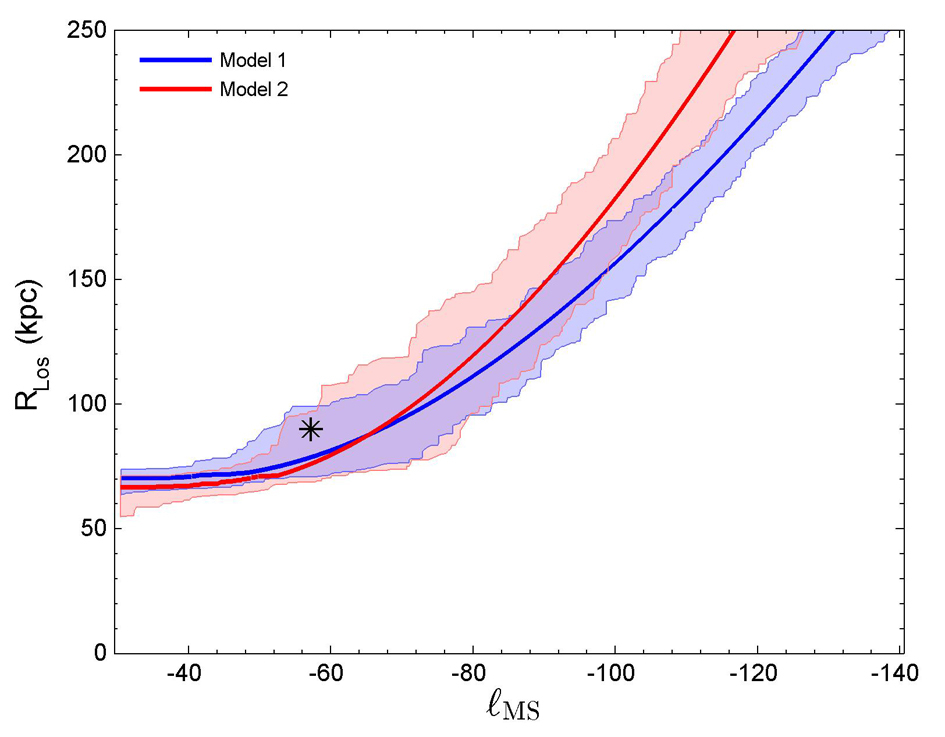}
\caption{Fit of the distance along the Stream as a function of $\ell_{\rm{MS}}$ for Model 1 (\emph{blue}) and Model 2 (\emph{red}). The shaded regions indicate the confidence interval, calculated from the bootstrap distribution in both models. The \emph{dashed line} shows the direction of the South Galactic Pole.}
\label{fig:DistanceStream}
\end{figure}

The precise distance of the Magellanic Stream has important implications for its fundamental parameters. For example, the Stream's total
gas mass is critically dependent on its distance \citep{2003ApJ...586..170P}. The detectability of
certain stellar populations also depends on the Stream's distance. No stars have been
detected which has hampered distance estimates to date.

One constraint on distance is provided by the geometrical method presented in \cite{2008MNRAS.383.1686J}. Using the data from \cite{2003ApJ...586..170P}, they found that the tip of the stream is at a distance of $75\,\rm{kpc}$. The top panels in Figures \ref{fig:BeslaVel} and \ref{fig:Model2Vel} show that in both models the simulated Stream is at a distance greater than the one expected. 
The discrepancies between the models and the observations are due to the absence of the ram pressure term in modelling the interaction between the gas within the Clouds and the hot halo gas of our Galaxy. The dense Galactic environments will have a strong influence on the inclination and the distance of the Stream. Introducing this interaction, by modelling the Milky Way as a dynamically live galaxy or in the form of a drag force component in the equation of motion, will improve the final shape and inclination of the simulated Stream, as well as the Leading Arm. 

The similar trend of the distance as a function of $\ell_{MS}$, shared by the two models (as shown in Figure \ref{fig:DistanceStream}) is particularly interesting. In this Figure, the solid lines describe the fit to the simulated Stream for Model 1 (\emph{red}) and Model 2 (\emph{blue}), in order to show the trend; while the shaded regions provide the error on the fit, obtained from the bootstrap distribution. Both models have a similar distance between $-80^{\circ}\le\ell_{MS}\le{-30^{\circ}}$, with equal distance at the position of the South Galactic Pole (\emph{black star}) of $~80\;\rm{kpc}$. For $\ell_{MS}<-80^{\circ}$, the increase of the distance is steeper for Model 2 than for Model 1. 

An accurate distance for the Stream also has a bearing on resolving the long-standing mystery of the Stream's 
high levels of ionisation over the SGP.
The presence of a bright $\rm{H}{\alpha}$ emission around the South Galactic Pole ($\ell_{\rm{MS}}=-57^{\circ}$) cannot be explained by a Galactic UV radiation field (stars, gas, etc.). In a recent paper, \cite{2013ApJ...778...58B} argue that the photoionization levels along the Stream are best explained by a Seyfert flare model, consistent with the most viable explanation of the \emph{Fermi} bubbles \citep{2012ApJ...756..181G}. \cite{2013ApJ...778...58B} define an ionisation cone emanating from Sgr A* aligned roughly with the South Galactic Pole (SGP) and gas clouds within the cone are lit up by a Seyfert flare approximately $2\,\rm{Myr}$ ago. 

The energetic details of the past explosion critically
depend on the distance of the Stream. A near-distance of about 50 kpc lowers the
required energetics to about 10 per cent of the maximum Eddington luminosity required by Sgr A$^\star$.
A greater distance of 100 kpc pushes up the required luminosity close to its maximum value 
\citep[see their Appendix A]{2013ApJ...778...58B}. For the smaller distance, a shock cascade
acting along the Stream could conceivably account for the observed H$\alpha$ emission. But
this model breaks down for the larger distance due to the lower halo coronal density \citep{2007ApJ...670L.109B}.%-------------------------------------------------------
%---CONCLUSION
%----------------------------------------
\section{Conclusion}
We present a new and novel technique for the study of the interaction between the Magellanic Clouds and the Milky Way. By combining a genetic algorithm with full N-body simulations, we are able to identify the orbit of the Magellanic Clouds based on a direct comparison between simulations and observations. Previous studies have constrained the orbital parameters of the Magellanic Clouds-Milky Way system \citep{ 2009ApJ...691.1807R,2012ApJ...750...36D}; but this is the first time that the parameter search has been done considering both Clouds as full N-body systems. During the parameter search, the Magellanic Clouds are represented by dark matter halo and a disc component with total mass equal to $2.43\times10^{10}\,\rm{M_{\odot}}$ for LMC and $0.63\times10^{10}\,\rm{M_{\odot}}$ for SMC. 

The Milky Way is modelled as a three component potential with a Herquist bulge, Miyamoto-Nagai disc and a Navarro, Frenk and White dark matter halo. The latter depends on three parameters: the virial mass, the virial radius and the concentration parameters. In this analysis, the virial mass and concentration are independent parameters, free to span  the range given in Table \ref{tab:GAparam}; while the virial radius of the dark matter halo is instead directly calculated from its virial mass. Although the dark matter halo has the strongest influence on the motion of the Clouds, the particular choice of disc and bulge parameters influences the value of the Milky Way circular velocity, a crucial parameter for the orbit of the Clouds (see equation \ref{eq:val}). Therefore, for each selected virial mass and concentration, the circular velocity at the position of the Sun is directly calculated using the rotation curve of the Milky Way. 

We provided two orbital scenarios for the Clouds, resulting from two different models of the Milky Way. As seen in Figure \ref{fig:OrbitInfo}, both models support more traditional orbits around the main Galaxy. This is not surprising, since traditional orbits are expected for a $10^{10}\,\rm{M_{\odot}}$ mass LMC, particularly with a high ($\sim$245 kms${^{-1}}$, in both models) circular velocity \citep{2012ApJ...759...99Z,2013ApJ...764..161K}. 
Interestingly, the values of the Milky Way parameters describe a less massive ($\le1.5\times10^{12}\,\rm{M_{\odot}}$) but more concentrated dark matter halo ($c>20$). We show that this is not odd, since studies of the kinematics of the Milky Way stellar halo also prefer such models, with higher concentration parameter than the one obtained in cosmological simulations \citep{2005MNRAS.364..433B,2012MNRAS.424L..44D}. 

 The orbits described in Figure \ref{fig:OrbitInfo} are selected by using the star formation history as the only condition on the LMC-SMC interaction. The two common starbursts, one 2-3 Gyr ago and the other 400 Myr ago, can be interpreted as evidence for two encounters between the Clouds \citep{2009AJ....138.1243H,2004AJ....127.1531H}. No other orbital criteria are applied, especially on the evolution around the Milky Way. It might be argued that using the star formation history as a constraint on the orbits introduces bias, as there are observational error related to the time of the star bursts However, the power of this method is that any constraints on the actual values can be introduced into the fitness function and, therefore, used for better selecting the orbits. 
 
 The solution of the GA suggests that not all of the LMC-SMC orbits are possible, but they depend on the eccentricity of the orbit on which SMC lies. Figure \ref{fig:F2vsEcc} confirms this dependency. The orbit of SMC around LMC needs to have an eccentricity between $0.6-0.7$, otherwise it will decay too quickly into the LMC or it will be pushed away by its interaction with the LMC halo. 

As a result  of the selected orbits, the middle and right panels in Figure \ref{fig:Zea} show the presence of an extended tail, a leading arm and a bridge of gas connecting the two galaxies.  Our models also support the presence of a stellar bridge, which formed after the last encounter. The stars in this region were tidally stripped from both Clouds at the formation epoch of the gaseous Bridge. The presence of the stars in the regions are also suggested by recent spectroscopic and photometric studies of the SMC and the Magellanic Bridge \citep{2013A&A...551A..78B,2013ApJ...779..145N,2014MNRAS.442.1680D}.  

The formation mechanism of the Stream is common in both models; the interactions between the Clouds lead to the formation of the Magellanic System \citep{2010ApJ...721L..97B,2012MNRAS.421.2109B}. The Clouds form a binary pair at least for the last $3\,\rm{Gyr}$ and the encounters between LMC and SMC are strong enough to strip material away from about $2\,\rm{Gyr}$ ago, mainly from the Small Cloud, in agreements with previous models \citep{2006MNRAS.371..108C,2012ApJ...750...36D} and the recent results from $\rm{HST/COS}$ and $\rm{VLT/UVES}$ \citep{2013ApJ...772..110F}. The models also offer a good description of the Stream kinematics, showing a gradient of the line-of-sight velocity along the stream \citep{ 2003ApJ...586..170P, 2010ApJ...723.1618N}. However, the tip the simulated streams do not fully reproduce the observation (see right panel in Figure \ref{fig:Zea}). As discussed in \cite{2012MNRAS.421.2109B}, the inclination of the SMC's disk with respect to the LMC-SMC orbital plane an important role in shaping the location of the Stream: for a fixed orbit, different orientation of the SMC disk can affect entirely the formation of the Magellanic Stream. Further investigations on the role of this parameter will allow to a better model of this system.

In addition, both models fail to reproduce the location of the Leading Arm and its bifurcation in full. It will be interesting to add the location of this feature as a condition to the genetic algorithm, in order to select the models based on their ability to reproduce the Leading Arm. 

A crucial step forward in the analysis of the interaction between the Magellanic System and the Milky Way will be a proper dynamical analysis of the main galaxy. By modelling the latter as a full N-body system, it will allow to study the effect on the Milky Way halo due to the Magellanic System (momentum conservation), which has never been carried out
in a statistical sense, but only in single high-resolution simulations.

 \section*{ACKNOWLEDGEMENTS}
M.~G. acknowledges the Australia Postgraduate Award (APA) for the support
of her PhD candidature.  G.~F.~L. thanks the Australian research council for ARC Future Fellowship (FT100100268).
The authors gratefully acknowledge fruitful conversations with Dr. Sanjib Sharma, Dr. Pascal Elahi, Dr. Nicholas Bate  and Foivos Diakogiannis.
Computational resources used in this work were provided by
the University of Sydney High Performance Facilities.
\bibliographystyle{mn2e}
\bibliography{GuglielmoMagellanicStream}

\begin{thebibliography}{85}
\expandafter\ifx\csname natexlab\endcsname\relax\def\natexlab#1{#1}\fi

\bibitem[{{Bagheri}, {Cioni} \& {Napiwotzki}(2013){Bagheri}, {Cioni}, \&
  {Napiwotzki}}]{2013A&A...551A..78B}
{Bagheri} G., {Cioni} M.-R.~L., {Napiwotzki} R., 2013, \aap, 551, A78

\bibitem[{{Barger}, {Haffner} \& {Bland-Hawthorn}(2013){Barger}, {Haffner}, \&
  {Bland-Hawthorn}}]{2013ApJ...771..132B}
{Barger} K.~A., {Haffner} L.~M., {Bland-Hawthorn} J., 2013, \apj, 771, 132

\bibitem[{{Battaglia} {et~al}\mbox{.}(2005){Battaglia}, {Helmi}, {Morrison},
  {Harding}, {Olszewski}, {Mateo}, {Freeman}, {Norris}, \&
  {Shectman}}]{2005MNRAS.364..433B}
{Battaglia} G. {et~al.}, 2005, \mnras, 364, 433

\bibitem[{{Bekki} \& {Chiba}(2005)}]{2005MNRAS.356..680B}
{Bekki} K., {Chiba} M., 2005, \mnras, 356, 680

\bibitem[{{Besla} {et~al}\mbox{.}(2007){Besla}, {Kallivayalil}, {Hernquist},
  {Robertson}, {Cox}, {van der Marel}, \& {Alcock}}]{2007ApJ...668..949B}
{Besla} G., {Kallivayalil} N., {Hernquist} L., {Robertson} B., {Cox} T.~J.,
  {van der Marel} R.~P., {Alcock} C., 2007, \apj, 668, 949

\bibitem[{{Besla} {et~al}\mbox{.}(2010){Besla}, {Kallivayalil}, {Hernquist},
  {van der Marel}, {Cox}, \& {Kere{\v s}}}]{2010ApJ...721L..97B}
{Besla} G., {Kallivayalil} N., {Hernquist} L., {van der Marel} R.~P., {Cox}
  T.~J., {Kere{\v s}} D., 2010, \apjl, 721, L97

\bibitem[{{Besla} {et~al}\mbox{.}(2012){Besla}, {Kallivayalil}, {Hernquist},
  {van der Marel}, {Cox}, \& {Kerevs}}]{2012MNRAS.421.2109B}
{Besla} G., {Kallivayalil} N., {Hernquist} L., {van der Marel} R.~P., {Cox}
  T.~J., {Kerevs} D., 2012, \mnras, 421, 2109

\bibitem[{{Binney} \& {Tremaine}(2008)}]{2008gady.book.....B}
{Binney} J., {Tremaine} S., 2008, {Galactic Dynamics: Second Edition}.
  Princeton University Press

\bibitem[{{Bland-Hawthorn} {et~al}\mbox{.}(2013){Bland-Hawthorn}, {Maloney},
  {Sutherland}, \& {Madsen}}]{2013ApJ...778...58B}
{Bland-Hawthorn} J., {Maloney} P.~R., {Sutherland} R.~S., {Madsen} G.~J., 2013,
  \apj, 778, 58

\bibitem[{{Bland-Hawthorn} {et~al}\mbox{.}(2007){Bland-Hawthorn}, {Sutherland},
  {Agertz}, \& {Moore}}]{2007ApJ...670L.109B}
{Bland-Hawthorn} J., {Sutherland} R., {Agertz} O., {Moore} B., 2007, \apjl,
  670, L109

\bibitem[{{Boylan-Kolchin}, {Besla} \& {Hernquist}(2011){Boylan-Kolchin},
  {Besla}, \& {Hernquist}}]{2011MNRAS.414.1560B}
{Boylan-Kolchin} M., {Besla} G., {Hernquist} L., 2011, \mnras, 414, 1560

\bibitem[{{Boylan-Kolchin} {et~al}\mbox{.}(2013){Boylan-Kolchin}, {Bullock},
  {Sohn}, {Besla}, \& {van der Marel}}]{2013ApJ...768..140B}
{Boylan-Kolchin} M., {Bullock} J.~S., {Sohn} S.~T., {Besla} G., {van der Marel}
  R.~P., 2013, \apj, 768, 140

\bibitem[{{Brewer} \& {Lewis}(2005)}]{2005PASA...22..128B}
{Brewer} B.~J., {Lewis} G.~F., 2005, \pasa, 22, 128

\bibitem[{{Br{\"u}ns} {et~al}\mbox{.}(2005){Br{\"u}ns}, {Kerp},
  {Staveley-Smith}, {Mebold}, {Putman}, {Haynes}, {Kalberla}, {Muller}, \&
  {Filipovic}}]{2005A&A...432...45B}
{Br{\"u}ns} C. {et~al.}, 2005, \aap, 432, 45

\bibitem[{{Bullock} \& {Johnston}(2005)}]{2005ApJ...635..931B}
{Bullock} J.~S., {Johnston} K.~V., 2005, \apj, 635, 931

\bibitem[{{Charbonneau}(1995)}]{1995ApJS..101..309C}
{Charbonneau} P., 1995, \apjs, 101, 309

\bibitem[{{Cioni} {et~al}\mbox{.}(2000){Cioni}, {van der Marel}, {Loup}, \&
  {Habing}}]{2000AA...359..601C}
{Cioni} M.-R.~L., {van der Marel} R.~P., {Loup} C., {Habing} H.~J., 2000, \aap,
  359, 601

\bibitem[{{Connors}, {Kawata} \& {Gibson}(2006){Connors}, {Kawata}, \&
  {Gibson}}]{2006MNRAS.371..108C}
{Connors} T.~W., {Kawata} D., {Gibson} B.~K., 2006, \mnras, 371, 108

\bibitem[{{Costa} {et~al}\mbox{.}(2009){Costa}, {M{\'e}ndez}, {Pedreros},
  {Moyano}, {Gallart}, {No{\"e}l}, {Baume}, \& {Carraro}}]{2009AJ....137.4339C}
{Costa} E., {M{\'e}ndez} R.~A., {Pedreros} M.~H., {Moyano} M., {Gallart} C.,
  {No{\"e}l} N., {Baume} G., {Carraro} G., 2009, \aj, 137, 4339

\bibitem[{{Deason} {et~al}\mbox{.}(2012){Deason}, {Belokurov}, {Evans}, \&
  {An}}]{2012MNRAS.424L..44D}
{Deason} A.~J., {Belokurov} V., {Evans} N.~W., {An} J., 2012, \mnras, 424, L44

\bibitem[{{Diaz} \& {Bekki}(2012)}]{2012ApJ...750...36D}
{Diaz} J.~D., {Bekki} K., 2012, \apj, 750, 36

\bibitem[{{Dobbie} {et~al}\mbox{.}(2014){Dobbie}, {Cole}, {Subramaniam}, \&
  {Keller}}]{2014MNRAS.442.1680D}
{Dobbie} P.~D., {Cole} A.~A., {Subramaniam} A., {Keller} S., 2014, \mnras, 442,
  1680

\bibitem[{{Dubois} {et~al}\mbox{.}(2012){Dubois}, {Pichon}, {Haehnelt}, {Kimm},
  {Slyz}, {Devriendt}, \& {Pogosyan}}]{2012MNRAS.423.3616D}
{Dubois} Y., {Pichon} C., {Haehnelt} M., {Kimm} T., {Slyz} A., {Devriendt} J.,
  {Pogosyan} D., 2012, \mnras, 423, 3616

\bibitem[{{Fox} {et~al}\mbox{.}(2013){Fox}, {Richter}, {Wakker}, {Lehner},
  {Howk}, {Ben Bekhti}, {Bland-Hawthorn}, \& {Lucas}}]{2013ApJ...772..110F}
{Fox} A.~J., {Richter} P., {Wakker} B.~P., {Lehner} N., {Howk} J.~C., {Ben
  Bekhti} N., {Bland-Hawthorn} J., {Lucas} S., 2013, \apj, 772, 110

\bibitem[{{Fox} {et~al}\mbox{.}(2014){Fox}, {Wakker}, {Barger}, {Hernandez},
  {Richter}, {Lehner}, {Bland-Hawthorn}, {Charlton}, {Westmeier}, {Thom},
  {Tumlinson}, {Misawa}, {Howk}, {Haffner}, {Ely}, {Rodriguez-Hidalgo}, \&
  {Kumari}}]{2014arXiv1404.5514F}
{Fox} A.~J. {et~al.}, 2014, \apj, 787, 147

\bibitem[{{Gardiner}, {Sawa} \& {Fujimoto}(1994){Gardiner}, {Sawa}, \&
  {Fujimoto}}]{1994MNRAS.266..567G}
{Gardiner} L.~T., {Sawa} T., {Fujimoto} M., 1994, \mnras, 266, 567

\bibitem[{{Gnedin} {et~al}\mbox{.}(2004){Gnedin}, {Kravtsov}, {Klypin}, \&
  {Nagai}}]{2004ApJ...616...16G}
{Gnedin} O.~Y., {Kravtsov} A.~V., {Klypin} A.~A., {Nagai} D., 2004, \apj, 616,
  16

\bibitem[{{Guo} \& {Mathews}(2012)}]{2012ApJ...756..181G}
{Guo} F., {Mathews} W.~G., 2012, \apj, 756, 181

\bibitem[{{Harris} \& {Zaritsky}(2004)}]{2004AJ....127.1531H}
{Harris} J., {Zaritsky} D., 2004, \aj, 127, 1531

\bibitem[{{Harris} \& {Zaritsky}(2006)}]{2006AJ....131.2514H}
{Harris} J., {Zaritsky} D., 2006, \aj, 131, 2514

\bibitem[{{Harris} \& {Zaritsky}(2009)}]{2009AJ....138.1243H}
{Harris} J., {Zaritsky} D., 2009, \aj, 138, 1243

\bibitem[{{Hashimoto}, {Funato} \& {Makino}(2003){Hashimoto}, {Funato}, \&
  {Makino}}]{2003ApJ...582..196H}
{Hashimoto} Y., {Funato} Y., {Makino} J., 2003, \apj, 582, 196

\bibitem[{{Hernquist}(1990)}]{1990ApJ...356..359H}
{Hernquist} L., 1990, \apj, 356, 359

\bibitem[{{Irwin}, {Demers} \& {Kunkel}(1990){Irwin}, {Demers}, \&
  {Kunkel}}]{1990AJ.....99..191I}
{Irwin} M.~J., {Demers} S., {Kunkel} W.~E., 1990, \aj, 99, 191

\bibitem[{{James} \& {Ivory}(2011)}]{2011MNRAS.411..495J}
{James} P.~A., {Ivory} C.~F., 2011, \mnras, 411, 495

\bibitem[{{Jin} \& {Lynden-Bell}(2008)}]{2008MNRAS.383.1686J}
{Jin} S., {Lynden-Bell} D., 2008, \mnras, 383, 1686

\bibitem[{{Kafle} {et~al}\mbox{.}(2014){Kafle}, {Sharma}, {Lewis}, \&
  {Bland-Hawthorn}}]{2014Kafle}
{Kafle} P., {Sharma} S., {Lewis} G.~F., {Bland-Hawthorn} J., 2014, An
  cbservational constrained 3d model of the milky way potential, submitted
  {\apj}

\bibitem[{{Kafle} {et~al}\mbox{.}(2012){Kafle}, {Sharma}, {Lewis}, \&
  {Bland-Hawthorn}}]{2012ApJ...761...98K}
{Kafle} P.~R., {Sharma} S., {Lewis} G.~F., {Bland-Hawthorn} J., 2012, \apj,
  761, 98

\bibitem[{{Kallivayalil}, {van der Marel} \& {Alcock}(2006){Kallivayalil}, {van
  der Marel}, \& {Alcock}}]{2006ApJ...652.1213K}
{Kallivayalil} N., {van der Marel} R.~P., {Alcock} C., 2006, \apj, 652, 1213

\bibitem[{{Kallivayalil} {et~al}\mbox{.}(2013){Kallivayalil}, {van der Marel},
  {Besla}, {Anderson}, \& {Alcock}}]{2013ApJ...764..161K}
{Kallivayalil} N., {van der Marel} R.~P., {Besla} G., {Anderson} J., {Alcock}
  C., 2013, \apj, 764, 161

\bibitem[{{Klypin}, {Zhao} \& {Somerville}(2002){Klypin}, {Zhao}, \&
  {Somerville}}]{2002ApJ...573..597K}
{Klypin} A., {Zhao} H., {Somerville} R.~S., 2002, \apj, 573, 597

\bibitem[{{Larson}, {Tinsley} \& {Caldwell}(1980){Larson}, {Tinsley}, \&
  {Caldwell}}]{1980ApJ...237..692L}
{Larson} R.~B., {Tinsley} B.~M., {Caldwell} C.~N., 1980, \apj, 237, 692

\bibitem[{{Lin} \& {Lynden-Bell}(1977)}]{1977MNRAS.181...59L}
{Lin} D.~N.~C., {Lynden-Bell} D., 1977, \mnras, 181, 59

\bibitem[{{Macci{\`o}}, {Dutton} \& {van den Bosch}(2008){Macci{\`o}},
  {Dutton}, \& {van den Bosch}}]{2008MNRAS.391.1940M}
{Macci{\`o}} A.~V., {Dutton} A.~A., {van den Bosch} F.~C., 2008, \mnras, 391,
  1940

\bibitem[{{Mastropietro} {et~al}\mbox{.}(2005){Mastropietro}, {Moore}, {Mayer},
  {Wadsley}, \& {Stadel}}]{2005MNRAS.363..509M}
{Mastropietro} C., {Moore} B., {Mayer} L., {Wadsley} J., {Stadel} J., 2005,
  \mnras, 363, 509

\bibitem[{{McCumber}, {Garnett} \& {Dufour}(2005){McCumber}, {Garnett}, \&
  {Dufour}}]{2005AJ....130.1083M}
{McCumber} M.~P., {Garnett} D.~R., {Dufour} R.~J., 2005, \aj, 130, 1083

\bibitem[{{McMillan}(2011)}]{2011MNRAS.414.2446M}
{McMillan} P.~J., 2011, \mnras, 414, 2446

\bibitem[{{McMillan} \& {Binney}(2010)}]{2010MNRAS.402..934M}
{McMillan} P.~J., {Binney} J.~J., 2010, \mnras, 402, 934

\bibitem[{{Miyamoto} \& {Nagai}(1975)}]{1975PASJ...27..533M}
{Miyamoto} M., {Nagai} R., 1975, \pasj, 27, 533

\bibitem[{{Mo}, {Mao} \& {White}(1998){Mo}, {Mao}, \&
  {White}}]{1998MNRAS.295..319M}
{Mo} H.~J., {Mao} S., {White} S.~D.~M., 1998, \mnras, 295, 319

\bibitem[{{Moore} \& {Davis}(1994)}]{1994MNRAS.270..209M}
{Moore} B., {Davis} M., 1994, \mnras, 270, 209

\bibitem[{{Murai} \& {Fujimoto}(1980)}]{1980PASJ...32..581M}
{Murai} T., {Fujimoto} M., 1980, \pasj, 32, 581

\bibitem[{{Navarro}, {Frenk} \& {White}(1997){Navarro}, {Frenk}, \&
  {White}}]{1997ApJ...490..493N}
{Navarro} J.~F., {Frenk} C.~S., {White} S.~D.~M., 1997, \apj, 490, 493

\bibitem[{{Nichols} {et~al}\mbox{.}(2011){Nichols}, {Colless}, {Colless}, \&
  {Bland-Hawthorn}}]{2011ApJ...742..110N}
{Nichols} M., {Colless} J., {Colless} M., {Bland-Hawthorn} J., 2011, \apj, 742,
  110

\bibitem[{{Nidever}, {Majewski} \& {Burton}(2008){Nidever}, {Majewski}, \&
  {Burton}}]{2008ApJ...679..432N}
{Nidever} D.~L., {Majewski} S.~R., {Burton} W.~B., 2008, \apj, 679, 432

\bibitem[{{Nidever} {et~al}\mbox{.}(2010){Nidever}, {Majewski}, {Butler
  Burton}, \& {Nigra}}]{2010ApJ...723.1618N}
{Nidever} D.~L., {Majewski} S.~R., {Butler Burton} W., {Nigra} L., 2010, \apj,
  723, 1618

\bibitem[{{Nidever} {et~al}\mbox{.}(2013){Nidever}, {Monachesi}, {Bell},
  {Majewski}, {Mu{\~n}oz}, \& {Beaton}}]{2013ApJ...779..145N}
{Nidever} D.~L., {Monachesi} A., {Bell} E.~F., {Majewski} S.~R., {Mu{\~n}oz}
  R.~R., {Beaton} R.~L., 2013, \apj, 779, 145

\bibitem[{{No{\"e}l} {et~al}\mbox{.}(2009){No{\"e}l}, {Aparicio}, {Gallart},
  {Hidalgo}, {Costa}, \& {M{\'e}ndez}}]{2009ApJ...705.1260N}
{No{\"e}l} N.~E.~D., {Aparicio} A., {Gallart} C., {Hidalgo} S.~L., {Costa} E.,
  {M{\'e}ndez} R.~A., 2009, \apj, 705, 1260

\bibitem[{{Offer} \& {Bland-Hawthorn}(1998)}]{1998MNRAS.299..176O}
{Offer} A.~R., {Bland-Hawthorn} J., 1998, \mnras, 299, 176

\bibitem[{{Piatti} {et~al}\mbox{.}(2005){Piatti}, {Sarajedini}, {Geisler},
  {Seguel}, \& {Clark}}]{2005MNRAS.358.1215P}
{Piatti} A.~E., {Sarajedini} A., {Geisler} D., {Seguel} J., {Clark} D., 2005,
  \mnras, 358, 1215

\bibitem[{{Putman}, {Peek} \& {Joung}(2012){Putman}, {Peek}, \&
  {Joung}}]{2012ARA&A..50..491P}
{Putman} M.~E., {Peek} J.~E.~G., {Joung} M.~R., 2012, \araa, 50, 491

\bibitem[{{Putman} {et~al}\mbox{.}(2003){Putman}, {Staveley-Smith}, {Freeman},
  {Gibson}, \& {Barnes}}]{2003ApJ...586..170P}
{Putman} M.~E., {Staveley-Smith} L., {Freeman} K.~C., {Gibson} B.~K., {Barnes}
  D.~G., 2003, \apj, 586, 170

\bibitem[{{Rashkov} {et~al}\mbox{.}(2013){Rashkov}, {Pillepich}, {Deason},
  {Madau}, {Rockosi}, {Guedes}, \& {Mayer}}]{2013ApJ...773L..32R}
{Rashkov} V., {Pillepich} A., {Deason} A.~J., {Madau} P., {Rockosi} C.~M.,
  {Guedes} J., {Mayer} L., 2013, \apjl, 773, L32

\bibitem[{{Reid} {et~al}\mbox{.}(2009){Reid}, {Menten}, {Zheng}, {Brunthaler},
  {Moscadelli}, {Xu}, {Zhang}, {Sato}, {Honma}, {Hirota}, {Hachisuka}, {Choi},
  {Moellenbrock}, \& {Bartkiewicz}}]{2009ApJ...700..137R}
{Reid} M.~J. {et~al.}, 2009, \apj, 700, 137

\bibitem[{Robitaille \& Whitney(2009)}]{IAU:7916176}
Robitaille T.~P., Whitney B.~A., 2009, Proceedings of the International
  Astronomical Union, 5, 799

\bibitem[{{Robotham} {et~al}\mbox{.}(2012){Robotham}, {Baldry},
  {Bland-Hawthorn}, {Driver}, {Loveday}, {Norberg}, {Bauer}, {Bekki}, {Brough},
  {Brown}, {Graham}, {Hopkins}, {Phillipps}, {Power}, {Sansom}, \&
  {Staveley-Smith}}]{2012MNRAS.424.1448R}
{Robotham} A.~S.~G. {et~al.}, 2012, \mnras, 424, 1448

\bibitem[{{Ruzicka}, {Palous} \& {Theis}(2007){Ruzicka}, {Palous}, \&
  {Theis}}]{2007A&A...461..155R}
{Ruzicka} A., {Palous} J., {Theis} C., 2007, \aap, 461, 155

\bibitem[{{Ruzicka}, {Theis} \& {Palous}(2009){Ruzicka}, {Theis}, \&
  {Palous}}]{2009ApJ...691.1807R}
{Ruzicka} A., {Theis} C., {Palous} J., 2009, \apj, 691, 1807

\bibitem[{{Ruzicka}, {Theis} \& {Palous}(2010){Ruzicka}, {Theis}, \&
  {Palous}}]{2010ApJ...725..369R}
{Ruzicka} A., {Theis} C., {Palous} J., 2010, \apj, 725, 369

\bibitem[{{Sch{\"o}nrich}, {Binney} \& {Dehnen}(2010){Sch{\"o}nrich}, {Binney},
  \& {Dehnen}}]{2010MNRAS.403.1829S}
{Sch{\"o}nrich} R., {Binney} J., {Dehnen} W., 2010, \mnras, 403, 1829

\bibitem[{{Shattow} \& {Loeb}(2009)}]{2009MNRAS.392L..21S}
{Shattow} G., {Loeb} A., 2009, \mnras, 392, L21

\bibitem[{{Sohn} {et~al}\mbox{.}(2013){Sohn}, {Besla}, {van der Marel},
  {Boylan-Kolchin}, {Majewski}, \& {Bullock}}]{2013ApJ...768..139S}
{Sohn} S.~T., {Besla} G., {van der Marel} R.~P., {Boylan-Kolchin} M.,
  {Majewski} S.~R., {Bullock} J.~S., 2013, \apj, 768, 139

\bibitem[{{Springel}(2005)}]{2005MNRAS.364.1105S}
{Springel} V., 2005, \mnras, 364, 1105

\bibitem[{{Stanimirovi{\'c}}, {Staveley-Smith} \&
  {Jones}(2004){Stanimirovi{\'c}}, {Staveley-Smith}, \&
  {Jones}}]{2004ApJ...604..176S}
{Stanimirovi{\'c}} S., {Staveley-Smith} L., {Jones} P.~A., 2004, \apj, 604, 176

\bibitem[{{Theis}(1999)}]{1999RvMA...12..309T}
{Theis} C., 1999, in Reviews in Modern Astronomy, Vol.~12, Reviews in Modern
  Astronomy, {Schielicke} R.~E., ed., p. 309

\bibitem[{{Toomre} \& {Toomre}(1972)}]{1972ApJ...178..623T}
{Toomre} A., {Toomre} J., 1972, \apj, 178, 623

\bibitem[{{van der Marel} {et~al}\mbox{.}(2002){van der Marel}, {Alves},
  {Hardy}, \& {Suntzeff}}]{2002AJ....124.2639V}
{van der Marel} R.~P., {Alves} D.~R., {Hardy} E., {Suntzeff} N.~B., 2002, \aj,
  124, 2639

\bibitem[{{van der Marel} \& {Kallivayalil}(2014)}]{2014ApJ...781..121V}
{van der Marel} R.~P., {Kallivayalil} N., 2014, \apj, 781, 121

\bibitem[{{Vieira} {et~al}\mbox{.}(2010){Vieira}, {Girard}, {van Altena},
  {Zacharias}, {Casetti-Dinescu}, {Korchagin}, {Platais}, {Monet}, {L{\'o}pez},
  {Herrera}, \& {Castillo}}]{2010AJ....140.1934V}
{Vieira} K. {et~al.}, 2010, \aj, 140, 1934

\bibitem[{{Vogelsberger} {et~al}\mbox{.}(2013){Vogelsberger}, {Genel},
  {Sijacki}, {Torrey}, {Springel}, \& {Hernquist}}]{2013MNRAS.436.3031V}
{Vogelsberger} M., {Genel} S., {Sijacki} D., {Torrey} P., {Springel} V.,
  {Hernquist} L., 2013, \mnras, 436, 3031

\bibitem[{{Wahde}(1998)}]{1998A&AS..132..417W}
{Wahde} M., 1998, \aaps, 132, 417

\bibitem[{{Widrow}, {Pym} \& {Dubinski}(2008){Widrow}, {Pym}, \&
  {Dubinski}}]{2008ApJ...679.1239W}
{Widrow} L.~M., {Pym} B., {Dubinski} J., 2008, \apj, 679, 1239

\bibitem[{{Xue} {et~al}\mbox{.}(2008){Xue}, {Rix}, {Zhao}, {Re Fiorentin},
  {Naab}, {Steinmetz}, {van den Bosch}, {Beers}, {Lee}, {Bell}, {Rockosi},
  {Yanny}, {Newberg}, {Wilhelm}, {Kang}, {Smith}, \&
  {Schneider}}]{2008ApJ...684.1143X}
{Xue} X.~X. {et~al.}, 2008, \apj, 684, 1143

\bibitem[{{Zentner} \& {Bullock}(2003)}]{2003ApJ...598...49Z}
{Zentner} A.~R., {Bullock} J.~S., 2003, \apj, 598, 49

\bibitem[{{Zhang} {et~al}\mbox{.}(2012){Zhang}, {Lin}, {Burkert}, \&
  {Oser}}]{2012ApJ...759...99Z}
{Zhang} X., {Lin} D.~N.~C., {Burkert} A., {Oser} L., 2012, \apj, 759, 99

\end{thebibliography}

\end{document}